\documentclass[preprints,article,accept,moreauthors,pdftex]{mdpi}

\firstpage{1} 
\pubvolume{1}
\issuenum{1}
\articlenumber{0}
\pubyear{2021}
\copyrightyear{2021}
\datereceived{} 
\dateaccepted{} 
\datepublished{} 
\hreflink{https://doi.org/} 

\usepackage{bm}
\usepackage{bbm}
\usepackage{upgreek} 
\usepackage{amsmath}
\usepackage{graphicx}
\usepackage{tikz} 
\usetikzlibrary{shapes.geometric, arrows} 
\usepackage{caption}
\usepackage{subcaption}
\usepackage{verbatim}
\usepackage{hyperref} 

\Title{A unified Abaqus implementation of the phase field fracture method using only a user material subroutine}

\TitleCitation{A unified implementation of the phase field fracture method}

\Author{Yousef Navidtehrani $^{1}$, Covadonga Beteg\'{o}n $^{1}$ and Emilio Mart\'{\i}nez-Pa\~neda $^{2,}$*}

\AuthorNames{Yousef Navidtehrani, Covadonga Beteg\'{o}n and Emilio Mart\'{\i}nez-Pa\~neda}

\AuthorCitation{Navidtehrani, Y.; Beteg\'{o}n, C.; Mart\'{\i}nez-Pa\~neda, E.}

\address{%
$^{1}$ \quad Department of Construction and Manufacturing Engineering, University of Oviedo, Gij\'{o}n 33203, Spain\\
$^{2}$ \quad Department of Civil and Environmental Engineering, Imperial College London, London SW7 2AZ, UK}

\corres{Correspondence: e.martinez-paneda@imperial.ac.uk}

\abstract{We present a simple and robust implementation of the phase field fracture method in Abaqus. Unlike previous works, only a user material (UMAT) subroutine is used. This is achieved by exploiting the analogy between the phase field balance equation and heat transfer, which avoids the need for a user element mesh and enables taking advantage of Abaqus' in-built features. A unified theoretical framework and its implementation are presented, suitable for any arbitrary choice of crack density function and fracture driving force. Specifically, the framework is exemplified with the so-called \texttt{AT1}, \texttt{AT2} and phase field-cohesive zone models (\texttt{PF-CZM}). Both staggered and monolithic solution schemes are handled. We demonstrate the potential and robustness of this new implementation by addressing several paradigmatic 2D and 3D boundary value problems. The numerical examples show how the current implementation can be used to reproduce numerical and experimental results from the literature, and efficiently capture advanced features such as complex crack trajectories, crack nucleation from arbitrary sites and contact problems. The code developed can be downloaded from \url{www.empaneda.com/codes}.}

\keyword{Abaqus; Phase field fracture; Finite element analysis; UMAT; fracture mechanics.} 

\begin{document}

\section{Introduction}

Variational phase field methods for fracture are enjoying a notable success \cite{Wu2020,PTRSA2021}. Among many others, applications include shape memory alloys \cite{CMAME2021}, glass laminates \cite{Freddi2020,Schmidt2020}, hydrogen-embrittled alloys \cite{CMAME2018,JMPS2020}, dynamic fracture \cite{Borden2012,McAuliffe2016}, fibre-reinforced composites \cite{Alessi2017,Quintanas-Corominas2019,Alessi2019,CST2021}, functionally graded materials \cite{CPB2019,Kumar2021,DT2020}, fatigue crack growth \cite{Lo2019,Carrara2020}, and masonry structures \cite{Freddi2011}. The key to the success of the phase field paradigm in fracture mechanics is arguably three-fold. Firstly, the phase field paradigm can override the computational challenges associated with direct tracking of the evolving solid-crack interface. The interface is made spatially diffuse by using an auxiliary variable, the phase field $\phi$, which varies smoothly between the solid and crack phases and evolves based on a suitable governing equation. Such a paradigm has also opened new horizons in the modelling of other interfacial problems such as microstructural evolution \cite{Provatas2011} or corrosion \cite{JMPS2021}. Secondly, phase field modelling has provided a suitable platform for the simple yet rigorous fracture thermodynamics principles first presented by Griffith \cite{Griffith1920}. This energy-based approach enables overcoming the issues associated with local approaches based on stress intensity factors, such as the need for \textit{ad hoc} criteria for determining the crack propagation direction \cite{Francfort1998,Bourdin2000}. Thirdly, phase field fracture modelling has shown to be very compelling and robust from a computational viewpoint. Advanced fracture features such as complex crack trajectories, crack branching, nucleation, and merging can be captured in arbitrary geometries and dimensions, and on the original finite element mesh (see, e.g., \cite{Borden2016,Miehe2016b,TAFM2020c,Wu2021}). Also, computations can be conducted in a Backward Euler setting without the convergence issues observed when using other computational fracture methods. One reason behind this robustness is the flexibility introduced by solving the phase field, a damage-like variable, independently from the deformation problem. So-called \emph{staggered} solution schemes have been presented to exploit this flexibility by computing sequentially the displacement and phase field solutions \cite{Miehe2010a}, avoiding computationally-demanding phenomena such as snap-backs.\\

The success of phase field modelling has been, not surprisingly, accompanied by a vast literature devoted to the development of open-source codes and finite element implementations of variational phase field methods for fracture. These works have been aimed at both commercial finite element packages such as COMSOL \cite{Zhou2018a} and open source platforms like FEniCS \cite{Hirshikesh2019b}. The development of phase field fracture implementations in the commercial package Abaqus has received particular attention \cite{Msekh2015,Liu2016a,Molnar2017,Fang2019,Molnar2020,Wu2020c,AES2021}, due to its popularity in the solid mechanics community. However, these works require the use of multiple user subroutines, most often including a user element (UEL) subroutine. Abaqus' in-built elements cannot be employed due to the need of solving for the phase field $\phi$ as a nodal degree-of-freedom (DOF). Having to adopt a user-defined finite element carries multiple limitations; namely, post-processing requires the use of a \emph{dummy} mesh or \textit{ad hoc} scripts, and the majority of in-built features of Abaqus cannot be exploited, as the software suite is effectively used as a solver. In this work, we overcome these limitations by presenting a new implementation that only requires the use of a user material (UMAT) subroutine. The simple yet robust implementation presented is achieved by taking advantage of the analogy between the phase field evolution equation and heat transfer. This not only greatly simplifies the use of Abaqus for conducting phase field fracture studies but also enables taking advantage of the many in-built features provided by this commercial package. In addition, we present a generalised theoretical and numerical framework that encapsulates what are arguably the three most popular phase field fracture models presented to date: (i) the so-called \texttt{AT2} model \cite{Bourdin2000}, based on the Ambrosio and Tortorelli regularisation of the Mumford-Shah functional \cite{Ambrosio1990}, (ii) the \texttt{AT1} model \cite{Pham2011}, which includes an elastic phase in the damage response, and (iii) the phase field-cohesive zone model \texttt{PF-CZM} \cite{Wu2017,Wu2018a}, aimed at providing an explicit connection to the material strength. Our framework also includes two strain energy decompositions to prevent damage in compressive states: the spectral split \cite{Miehe2010a} and the volumetric-deviatoric one \cite{Amor2009} - both available in the context of \emph{anisotropic} and \emph{hybrid} formulations \cite{Ambati2015}. Moreover, the implementation can use both \emph{monolithic} and \emph{staggered} solution schemes, enhancing its robustness. Two example codes are provided with this work (see \url{www.empaneda.com/codes}), both capable of handling 2D and 3D analyses without any modification. One is a simple 33-line code, which showcases the simplicity of this approach by adopting the most widely used constitutive choices (\texttt{AT2}, no split). The other one is an extended version, with all the features mentioned above, aimed at providing a unified implementation for phase field fracture. To the authors' knowledge, the present work provides the simplest Abaqus implementation of the phase field fracture method.\\

The remainder of this manuscript is organised as follows. In Section \ref{Sec:Theory} we provide a generalised formulation for phase field fracture, which can accommodate a myriad of constitutive choices. This is exemplified with the \texttt{AT2}, \texttt{AT1} and \texttt{CZ-PFM} models. Then, in Section \ref{Sec:FEM}, the details of the finite element implementation are presented, including the analogy with heat transfer and the particularities of the Abaqus usage. The potential of the implementation presented is showcased in Section \ref{Sec:Results}, where several boundary value problems of particular interest are addressed. Namely, (i) a three-point bending test, to compare with the results obtained with other numerical methods; (ii) a concrete single-edge notched beam, to compare with experimental data; (iii) a notched plate with a hole, to simulate complex crack paths, merging and nucleation; and (iv) a 3D gear, where cracking occurs due to contact between the teeth. Finally, concluding remarks are given in Section \ref{Sec:Conclusions}.
 
\section{A generalised formulation for phase field fracture}
\label{Sec:Theory}

In this section, we formulate our generalised formulation, suitable for arbitrary constitutive choices of crack density function and fracture driving force. Consider an elastic body occupying an arbitrary domain $\Omega \subset {\rm I\!R}^n$ $(n \in[1,2,3])$, with an external boundary $\partial \Omega\subset {\rm I\!R}^{n-1}$ with outwards unit normal $\mathbf{n}$.

\subsection{Kinematics}

The primary kinematic variables are the displacement field vector $\mathbf{u}$ and the damage phase field $\phi$. In this work, we limit our attention to small strains and isothermal conditions. Consequently, the strain tensor $\bm{\varepsilon}$ reads
\begin{equation}
    \bm{\varepsilon} = \frac{1}{2}\left(\nabla\mathbf{u}^T+\nabla\mathbf{u}\right) \, .
\end{equation}

The nucleation and growth of cracks are described by using a smooth continuous scalar \emph{phase field} $\phi \in [0;1]$. The phase field describes the degree of damage, being $\phi=0$ when the material point is in its intact state and $\phi=1$ when the material point is fully broken. Since $\phi$ is smooth and continuous, discrete cracks are represented in a diffuse manner. The smearing of cracks is controlled by a phase field length scale $\ell$. The aim of this diffuse representation is to introduce, over a discontinuous surface $\Gamma$, the following approximation of the fracture energy \cite{Bourdin2000}:
\begin{equation}
    \Phi=\int_{\Gamma} G_c \, \text{d}S \approx \int_\Omega G_c\gamma(\phi,\nabla\phi) \, \text{d}V, \hspace{1cm} \text{for } \ell\rightarrow 0,
\end{equation}

\noindent where $\gamma$ is the so-called crack surface density functional and $G_c$ is the material toughness \citep{Griffith1920,Irwin1956}. This approximation circumvents the need to track discrete crack surfaces, a well-known challenge in computational fracture mechanics.

\subsection{Principle of virtual work. Balance of forces}

Now, we shall derive the balance equations for the coupled deformation-fracture system using the principle of virtual work. The Cauchy stress $\bm{\sigma}$ is introduced, which is work conjugate to the strains $\bm{\varepsilon}$. Also, a traction $\mathbf{T}$ is defined on the boundary of the solid $\partial\Omega$, work conjugate to the displacements $\mathbf{u}$. Regarding fracture, we introduce a scalar stress-like quantity $\omega$, which is work conjugate to the phase field $\phi$, and a phase field micro-stress vector $\bm{\upxi}$ that is work conjugate to the gradient of the phase field $\nabla\phi$. The phase field is assumed to be driven solely by the solution to the displacement problem. Thus, no external traction is associated with $\phi$. In the absence of body forces, the principle of virtual work reads:
\begin{equation}
 \int_\Omega \big\{ \bm{\sigma}:\delta\bm{\varepsilon}  + \omega\delta\phi+\bm{\upxi} \cdot \delta \nabla \phi
    \big\} \, \text{d}V =  \int_{\partial \Omega} \left( \mathbf{T} \cdot \delta \mathbf{u} \right) \, \text{d}S
\end{equation}
\noindent where $\delta$ denotes a virtual quantity. This equation must hold for an arbitrary domain $\Omega$ and for any kinematically admissible variations of the virtual quantities. Thus, by application of the Gauss divergence theorem, the local force balances are given by: 
\begin{equation}
    \begin{split}
        &\nabla\cdot\bm{\sigma}=0  \\
        &\nabla\cdot\bm{\upxi}-\omega =0
    \end{split}\hspace{2cm} \text{in } \,\, \Omega,\label{eq:balance}
\end{equation}

\noindent with natural boundary conditions: 
\begin{equation}
    \begin{split}
        \bm{\sigma}\cdot\mathbf{n}=\mathbf{T} \\
         \bm{\upxi} \cdot \mathbf{n}=0 
    \end{split} \hspace{2cm} \text{on } \,\, \partial\Omega.\label{eq:balance_BC}
\end{equation}

\subsection{Constitutive theory}

The constitutive theory is presented in a generalised fashion, and the \texttt{AT1} \cite{Pham2011}, \texttt{AT2} \cite{Bourdin2000} and \texttt{PF-CZM} \cite{Wu2017,Wu2018a} models are then derived as special cases. The total potential energy of the solid reads,
\begin{equation}\label{eq:TotalPotentialEnergy0}
W \left( \bm{\varepsilon} \left( \mathbf{u} \right), \, \phi, \,  \nabla \phi \right) = \psi \left( \bm{\varepsilon} \left( \mathbf{u} \right), \, g \left( \phi \right) \right)  +  \varphi \left( \phi, \, \nabla \phi \right)
\end{equation}

\noindent where $\psi$ is the elastic strain energy density and $\varphi$ is the fracture energy density. The former diminishes with increasing damage through the degradation function $g \left( \phi \right)$, which must fulfill the following conditions:
\begin{equation}
g \left( 0 \right) =1 , \,\,\,\,\,\,\,\,\,\, g \left( 1 \right) =0 , \,\,\,\,\,\,\,\,\,\, g' \left( \phi \right) \leq 0 \,\,\, \text{for} \,\,\, 0 \leq \phi \leq 1 \, .
\end{equation}

We proceed to formulate the fracture energy density as,
\begin{align}
    \varphi \left( \phi, \, \nabla \phi \right) = G_c \gamma(\phi, \nabla\phi) = G_c \dfrac{1}{4c_w\ell}\left( w(\phi) + \ell^2 |\nabla\phi|^2\right) \, .
\end{align}

\noindent where $\ell$ is the phase field length scale and $w(\phi)$ is the geometric crack function. The latter must fulfill the following conditions:
\begin{equation}
w \left( 0 \right) =0 , \,\,\,\,\,\,\,\,\,\, w \left( 1 \right) =1 , \,\,\,\,\,\,\,\,\,\, w' \left( \phi \right) \geq 0 \,\,\, \text{for} \,\,\, 0 \leq \phi \leq 1 \, .
\end{equation}

\noindent Also, $c_w$ is a scaling constant, related to the so-called geometric crack function:

\begin{equation}
    c_w = \int_0^1\sqrt{w(\zeta)} \, \text{d}\zeta \, .
\end{equation}

Damage is driven by the elastic energy stored in the solid, as characterised by the undamaged elastic strain energy density $\psi_0$. To prevent cracking under compressive strain states, the driving force for fracture can be decomposed into active $\psi^+_0$ and inactive $\psi^-_0$ parts. Accordingly, the elastic strain energy density can be defined as \cite{Miehe2010}:
\begin{equation}
    \psi \left( \bm{\varepsilon} \left( \mathbf{u} \right), \, g \left( \phi \right) \right) =\psi^+ \left( \bm{\varepsilon} \left( \mathbf{u} \right), \phi \right) + \psi_0^-  \left( \bm{\varepsilon} \left( \mathbf{u} \right) \right) =g \left( \phi \right) \psi_0^+ \left( \bm{\varepsilon} \left( \mathbf{u} \right) \right) + \psi_0^-  \left( \bm{\varepsilon} \left( \mathbf{u} \right) \right)
\end{equation}

\noindent Also, damage is an irreversible process: $\dot{\phi} \geq 0$. To enforce irreversibility, a history field variable $\mathcal{H}$ is introduced, which must satisfy the Karush–Kuhn–Tucker (KKT) conditions:
\begin{equation}
     \psi_{0}^+ - \mathcal{H} \leq 0 \text{,} \hspace{8mm} \dot{\mathcal{H}} \geq 0 \text{,} \hspace{8mm} \dot{\mathcal{H}}(\psi_{0}^+-\mathcal{H})=0 \, \, .
    \centering
\end{equation}

\noindent Accordingly, for a current time $t$, over a total time $\tau$, the history field can be defined as,
\begin{equation}
    \mathcal{H} = \text{max}_{t \in [0, \tau]} \psi_0^+ \left( t \right) \, .
\end{equation}

\noindent Consequently, the total potential energy of the solid (\ref{eq:TotalPotentialEnergy0}) can be re-formulated as,
\begin{equation}\label{eq:Free_energy}
     W =  {g(\phi)} \mathcal{H}  + \frac{G_c}{4 c_w }  \left(\frac{1}{ \ell} {w(\phi)}+\ell |\nabla \phi|^2\right) 
\end{equation}

Now we proceed to derive, in a generalised fashion, the fracture micro-stress variables $\omega$ and $\bm{\upxi}$. The scalar micro-stress $\omega$ is defined as:
\begin{equation}\label{eq:consOmega}
    \omega = \dfrac{\partial W}{\partial\phi} = {g^{\prime}(\phi)} \mathcal{H} +\frac{G_c}{4c_w \ell}  w^{\prime}(\phi) \, ,
\end{equation}

\noindent while the phase field micro-stress vector $\bm{\upxi}$ reads,
\begin{equation}\label{eq:consXi}
    \bm{\upxi} = \dfrac{\partial W}{\partial\nabla\phi} = \frac{\ell}{2c_w} G_{\mathrm{c}} \nabla \phi \, .
\end{equation}

\noindent Inserting these into the phase field balance equation (\ref{eq:balance}b), one reaches the following phase field evolution law:
\begin{equation}\label{eq:PhaseFieldStrongForm}
  \frac{G_c}{2c_w}  \left( \frac{w^{\prime}(\phi)}{2 \ell} - \ell \nabla^2 \phi \right) + {g^{\prime}(\phi)} \mathcal{H} = 0  
\end{equation}

We shall now make specific constitutive choices, particularising the framework to the so-called \texttt{AT2}, \texttt{AT1} and \texttt{PF-CZM} models.\\

\noindent \textit{Degradation function} $g \left( \phi \right)$. Both \texttt{AT2} and \texttt{AT1} models were originally formulated using a quadratic degradation function:
\begin{equation}
    g \left( \phi \right) = \left( 1 - \phi \right)^2 + \kappa \, 
\end{equation}

\noindent where $\kappa$ is a small, positive-valued constant that is introduced to prevent ill-conditioning when $\phi=1$. A value of $\kappa=1 \times 10^{-7}$ is adopted throughout this work. Alternatively, the \texttt{PF-CZM} model typically uses the following degradation function,
\begin{equation}
     g \left( \phi \right) = \frac{(1-\phi)^d} {(1-\phi)^d+ a \phi (1 + b \phi}  \, ,
\end{equation}

\noindent with,
\begin{equation}
    a=\frac{4 E G_c}{\pi \ell f_t^2} ,
\end{equation}

\noindent where $E$ denotes Young's modulus and $f_t$ is the tensile strength of the material. The choices of $b$ and $d$ depend on the softening law employed. Two commonly used softening laws are the linear one, with $b=-0.5$ and $d=2$, and the exponential one, with $b=2^{(5/3)}-3$ and $d=2.5$.\\

\noindent \textit{Dissipation function}. The dissipation function is governed by the magnitude of $w$ and, consequently, $c_w$. For the \texttt{AT2} model: $w (\phi)=\phi^2$ and $c=1/2$. Since $w'(0)=0$, this choice implies a vanishing threshold for damage. An initial, damage-free linear elastic branch is introduced in the \texttt{AT1} model, with the choices $w (\phi)=\phi$ and $c=2/3$. Finally, in the \texttt{PF-CZM} case we have $w (\phi)=2\phi-\phi^2$ and $c=\pi/4$.\\

\noindent \textit{Fracture driving force} $\psi_0^+$. The variationally consistent approach, as proposed in the original \texttt{AT2} model, is often referred to as the \emph{isotropic} formulation:
\begin{equation}
    \psi^+_0 \left( \bm{\varepsilon} \right) = \frac{1}{2} \bm{\varepsilon} : \bm{C}_0 : \bm{\varepsilon} = \frac{1}{2} \lambda \text{tr}^2 \left( \bm{\varepsilon} \right) + \mu \text{tr} ( \bm{\varepsilon}^2 ) \, , \,\,\,\,\,\, \psi^-_0 \left( \bm{\varepsilon} \right) = 0 \, .
\end{equation}

\noindent where $\bm{C}_0$ is the undamaged elastic stiffness tensor and $\lambda$ and $\mu$ are the Lamé parameters. In the context of the \texttt{AT1} and \texttt{AT2} models, damage under compression is prevented by decomposing the strain energy density following typically two approaches. One is the so-called \emph{volumetric-deviatoric} split, proposed by Amor \textit{et al.} \cite{Amor2009} reads,
\begin{equation}
    \psi^+_0 \left( \bm{\varepsilon} \right) = \frac{1}{2} K \langle \text{tr} \left( \bm{\varepsilon} \right) \rangle^2_+ + \mu \left( \bm{\varepsilon}' : \bm{\varepsilon}' \right) \, , \,\,\,\,\,\, \psi^-_0 \left( \bm{\varepsilon} \right) = \frac{1}{2} K \langle \text{tr} \left( \bm{\varepsilon} \right) \rangle^2_- \, .
\end{equation}

\noindent where $K$ is the bulk modulus, $\langle a \rangle_{\pm}=\left( a \pm |a| \right)/2$, and $\bm{\varepsilon}'=\bm{\varepsilon}-\text{tr}\left( \bm{\varepsilon}\right) \bm{I}/3$. The second one is the so-called \emph{spectral} decomposition, proposed by Miehe \textit{et al.} \cite{Miehe2010}, which builds upon the spectral decomposition of the strain tensor $\bm{\varepsilon}^\pm=\sum_{a=1}^3 \langle \varepsilon_I \rangle_\pm \mathbf{n}_I \otimes \mathbf{n}_I$, with $\varepsilon_I$ and $\mathbf{n}_I$ being, respectively, the strain principal strains and principal strain directions (with $I=1,2,3$). The strain energy decomposition then reads \cite{Miehe2010}:
\begin{equation}
    \psi_0^\pm \left( \bm{\varepsilon} \right) =  \frac{1}{2} \lambda \langle \text{tr} \left( \bm{\varepsilon} \right) \rangle^2_\pm + \mu \text{tr} \left[ \left( \bm{\varepsilon}^\pm \right)^2 \right]
\end{equation}

The split can be applied not only to the phase field balance law but also to the balance of linear momentum. Considering the split only in the phase field balance (\ref{eq:PhaseFieldStrongForm}) is typically referred to as the \emph{hybrid} approach \cite{Ambati2015}. Alternatively, an \emph{anisotropic} formulation can be used, such that the damaged version of the stress tensor $\bm{\sigma}$ is computed as,
\begin{equation}\label{eq:Anisotropic}
    \bm{\sigma} \left( \mathbf{u}, \phi \right) = g \left( \phi \right) \frac{\partial \psi_0^+ \left( \bm{\varepsilon} \right)}{\partial \bm{\varepsilon}} + \frac{\partial \psi_0^- \left( \bm{\varepsilon} \right)}{\partial \bm{\varepsilon}} \, .
\end{equation}

\noindent On the other hand, in the \texttt{PF-CZM} model the driving force for fracture is defined as \cite{Wu2017}:
\begin{equation}
      \psi_0^+ = \frac{ \langle \sigma_{1} \rangle_+^2}{2 E},
\end{equation}     

\noindent with the other term of the split being given by,
\begin{equation}
      \psi_0^- =\frac{1}{2E} \Big [ \sigma_1 \langle \sigma_1 \rangle_{-} + \sigma_2^2 + \sigma_3^2 - 2 \nu \left( \sigma_2 \sigma_3 + \sigma_1 \sigma_3 + \sigma_1 \sigma_2 \right) \Big],
\end{equation}

\noindent where $\nu$ is Poisson's ratio and $\sigma_i$ are the principal stresses, with $\sigma_1$ being the maximum principal (undamaged) stress. The variational consistency is lost but the failure surface of concrete under dominant tension can be well captured \cite{Wu2017}. This formulation is only used with the \emph{hybrid} approach.\\

In addition, it is important to note that for the \texttt{AT1} and \texttt{PF-CZM} models there is a minimum value of the fracture driving force, which we denote as $\mathcal{H}_{min}$. This is needed as otherwise $\phi \leq 0$, as can be observed by setting $\phi=0$ and solving the balance equation (\ref{eq:PhaseFieldStrongForm}). The magnitude of $\mathcal{H}_{min}$ is then given by the solution of (\ref{eq:PhaseFieldStrongForm}) for $\mathcal{H}$ under $\phi=0$. For the \texttt{AT1} case: $\mathcal{H}_{min}=3G_c/(16 \ell)$; while for the \texttt{PF-CZM} model: $\mathcal{H}_{min}=2G_c/(\pi a \ell)=f_t^2/(2E)$.

\section{Finite element implementation}
\label{Sec:FEM}

We proceed to present our finite element model. The unified phase field fracture theory presented in Section \ref{Sec:Theory} is numerically implemented in Abaqus using only a user material (UMAT) subroutine; i.e., at the integration point level. This is achieved by taking advantage of the similitude between the heat transfer law and the Helmholtz-type phase field balance equation. The analogy between heat transfer and phase field fracture is described in Section \ref{subsec:HETVAL}, while the specific details of the Abaqus implementation are given in Section \ref{subsec:ABAQUS}. The present implementation does not require the coding of residual and stiffness matrix terms; however, these are provided in Appendix \ref{Sec:AppendixFEM} for completeness.

\subsection{Heat transfer analogy}
\label{subsec:HETVAL}

Consider a solid with thermal conductivity $k$, specific heat $c_p$ and density $\rho$. In the presence of a heat source $r$, the evolution of the temperature field $T$ in time $t$ is given by the following balance law:
\begin{equation}\label{eq:HeatTransfer1}
    k \nabla^2 T - \rho c_p \frac{\partial T}{\partial t} = -r \, ,
\end{equation}

\noindent Under steady-state conditions the $\partial T / \partial t$ term vanishes and Eq. (\ref{eq:HeatTransfer1}) is reduced to,
\begin{equation}\label{eq:HeatTransfer2}
    k \nabla^2 T = -r
\end{equation}

\noindent Now, rearrange the phase field evolution law (\ref{eq:PhaseFieldStrongForm}) as,
\begin{equation}\label{eq:HeatTransferPhase}
    \nabla^2 \phi = \frac{g' \left( \phi \right) \mathcal{H} 2 c_w}{\ell G_c} + \frac{w' (\phi)}{2 \ell^2} \, .
\end{equation}

\noindent Equations (\ref{eq:HeatTransfer2}) and (\ref{eq:HeatTransferPhase}) are analogous upon considering the temperature to be equivalent to the phase field $T \equiv \phi$, assuming a unit thermal conductivity $k=1$, and defining the following heat flux due to internal heat generation,
\begin{equation}\label{eq:r}
    r = -\frac{g' \left( \phi \right) \mathcal{H} 2 c_w}{\ell G_c} - \frac{w' (\phi)}{2 \ell^2} \, .
\end{equation}

\noindent Finally, we also define the rate of change of heat flux ($r$) with temperature ($T\equiv \phi$),
\begin{equation}\label{eq:r_phi}
    \frac{\partial r}{\partial \phi} = -\frac{g'' \left( \phi \right) \mathcal{H} 2 c_w}{\ell G_c} - \frac{w'' (\phi)}{2 \ell^2} \, ,
\end{equation}

\noindent as required for the computation of the Jacobian matrix.

\subsection{Abaqus particularities}
\label{subsec:ABAQUS}

The analogy between heat transfer and phase field fracture lays the grounds for a straightforward implementation of variational phase field fracture models in Abaqus. Only a user material (UMAT) subroutine is needed, as it is possible to define within the UMAT a volumetric heat generation source (\ref{eq:r}) and its variation with respect to the temperature (\ref{eq:r_phi}). It must be noted that a recent version of Abaqus should be used, as the UMAT volumetric heat generation option does not function properly for versions older than 2020. The alternative for versions 2019 or older is to combine the UMAT with a heat flux (HETVAL) subroutine \cite{AES2021}.\\ 

Abaqus' in-built displacement-temperature elements can be used, significantly facilitating model development. The same process as for a standard Abaqus model can be followed, with a few exceptions. The user should employ an analysis step of the type coupled temperature-displacement, with a steady state response. Also, one should define as material properties the thermal conductivity $k$, the density $\rho$ and the specific heat $c_p$, all of them with a value of unity. To avoid editing the UMAT subroutine, the mechanical and fracture properties are provided as mechanical constants in the user material definition. Also, one should define a zero temperature initial condition $T (t=0)=0 \,  \forall \, \bm{x}$. No other pre-processing or post-processing steps are needed, everything can be done within the Abaqus/CAE graphical user interface, and the phase field solution can be visualised by plotting the nodal solution temperature (NT11). Inside of the UMAT, the material Jacobian $\bm{C}_0$ and the Cauchy stress $\bm{\sigma}_0$ are computed from the strain tensor. The current (undamaged) stress-strain state is used to determine the driving force for fracture, $\mathcal{H}$. Both $\bm{C}_0$ and $\bm{\sigma}_0$ are degraded using the current value of the phase field $\phi$ (temperature), which is passed to the subroutine by Abaqus, such that $\bm{C}=g(\phi)\bm{C}_0$ and $\bm{\sigma}=g(\phi)\bm{\sigma}_0$. Finally, $\mathcal{H}$ and $\phi$ are used to compute $r$ (\ref{eq:r}) and $\partial r / \partial \phi$ (\ref{eq:r_phi}), defined as the volumetric heat generation and its derivative with respect to the temperature. In its simplest form, the code requires only 33 lines.\\ 
 
The implementation also accommodates both \emph{monolithic} and \emph{staggered} schemes, enabling convergence even in computationally demanding problems. We choose not to define the non-diagonal, coupling terms of the displacement-phase field stiffness matrix; i.e. $\bm{K}_{\bm{u}\phi}=\bm{K}_{\phi \bm{u}}=\bm{0}$. This makes the stiffness matrix symmetric. By default, Abaqus assumes a non-symmetric system for coupled displacement-temperature analyses but one can configure the solver to deal with a symmetric system by using the separated solution technique. The current values of the phase field (temperature) and displacement solutions are provided to the subroutine, so they can used to update the relevant variables ($\bm{C}_0$, $\bm{\sigma}$, $r$ and $\partial r / \partial \phi$), such that the deformation and fracture problems are solved in a simultaneous (monolithic) manner. Conversely, one can use solution dependent state variables (SDVs) to store and use the history field of the previous increment $\mathcal{H}_t$, effectively freezing its value during the iterative procedure taking place for the current load increment. This is known as a single-pass staggered solution scheme. While single-pass staggered schemes are very robust, unconditional stability no longer holds and one should conduct a sensitivity analysis to ensure that the load increments employed are sufficiently small. Robustness and unconditional stability can be achieved by using quasi-Newton methods \cite{Wu2020a,TAFM2020}, but such option is not currently available in Abaqus for coupled temperature-displacement analyses. Independently of the solution scheme, it is known that phase field fracture analyses can achieve convergence after a large number of iterations \cite{Gerasimov2016,TAFM2020}. Thus, the solution controls are modified to enable this (see the example input file provided in \url{www.empaneda.com/codes}.

\section{Results}
\label{Sec:Results}

We address several paradigmatic boundary value problems to showcase the various features of the implementation, as well as its robustness and potential. First, we use the \texttt{PF-CZM} model to simulate fracture in a three-point bending experiment and compare the results with those obtained by Wells and Sluys \cite{Wells2001} using an enriched cohesive zone model. Secondly, we model mixed-mode fracture in a concrete beam to compare the crack trajectories predicted by the \texttt{AT2} model to those observed experimentally \cite{Schalangen1993}. Thirdly, cracking in a mortar plate with an eccentric hole is simulated to benchmark our predictions with the numerical and experimental results of Ambati \textit{et al.} \cite{Ambati2015}. Finally, the \texttt{AT1} model is used in a 3D analysis of crack nucleation and growth resulting from the interaction between two gears.

\subsection{Three-point bending test}

First, we follow the work by Wells and Sluys \cite{Wells2001} and model the failure of a beam subjected to three-point bending. In their analysis, Wells and Sluys combined the concepts of cohesive zone modelling and partition of unity, using an exponential traction-separation law \cite{Wells2001}. To establish a direct comparison, we choose to adopt the so-called phase field cohesive zone model (\texttt{PF-CZM}) \cite{Wu2017,Wu2018a} using the exponential degradation function.\\ 

\begin{figure}[H]
    \centering
    \begin{subfigure}[H]{0.5\textwidth}
    \hspace{-.4 cm}
        \includegraphics[width=\textwidth]{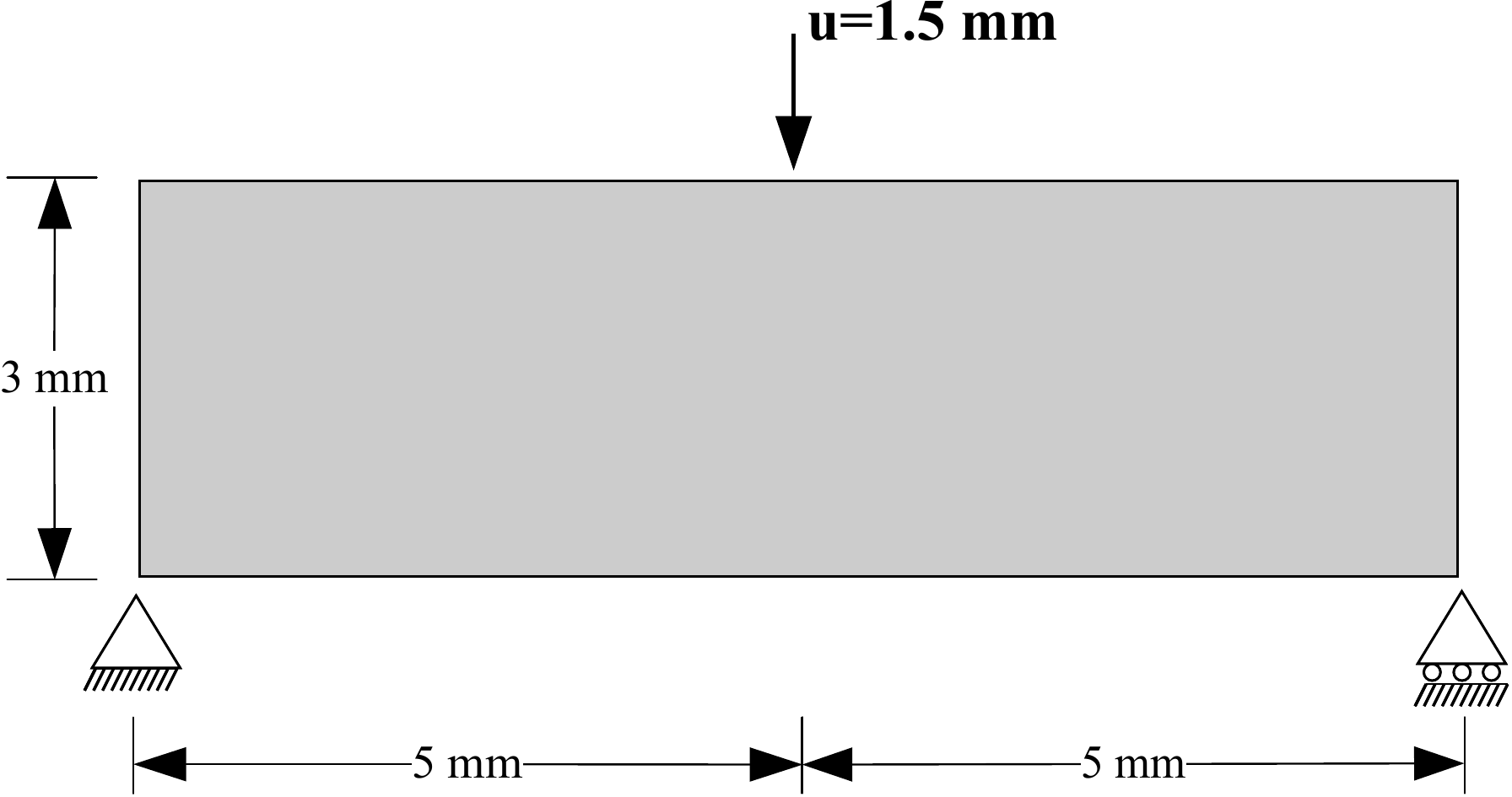}
        \caption{}
        \label{Fig:3point-geom}
    \end{subfigure}\vspace{0.02\textwidth}
    \begin{subfigure}[H]{0.45\textwidth}
        \includegraphics[width=\textwidth]{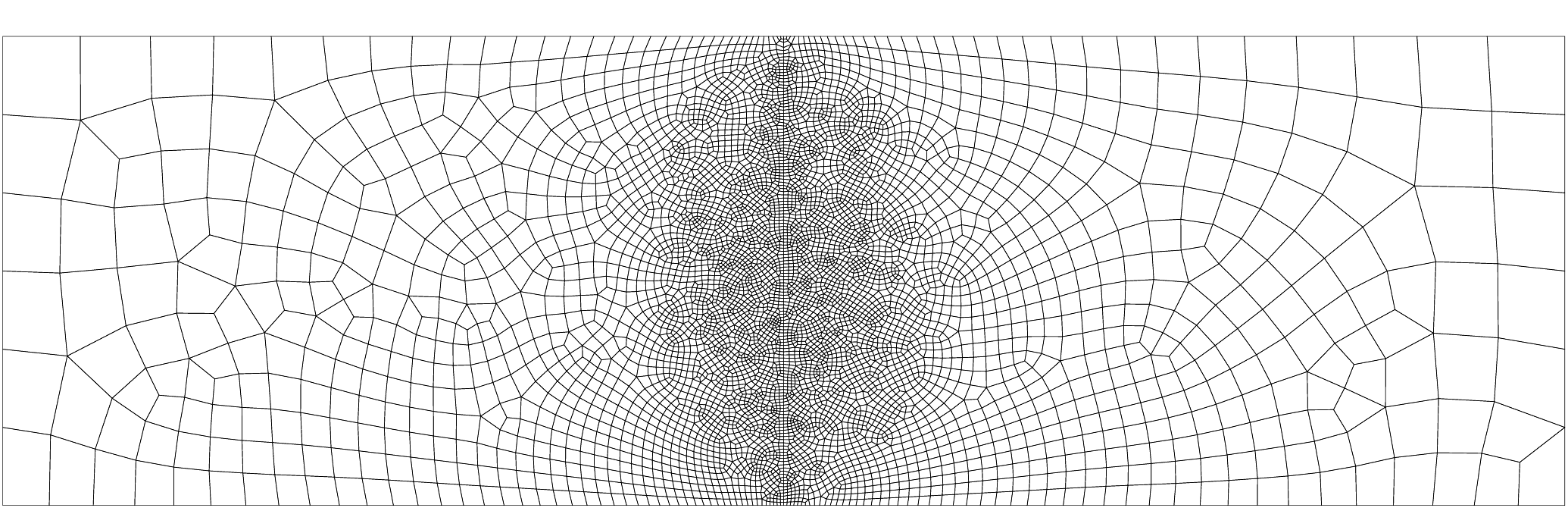}
        \caption{}
        \label{Fig:3point-mesh}
    \end{subfigure}\vspace{0.02\textwidth}
    \begin{subfigure}[H]{0.45\textwidth}
        \includegraphics[width=\textwidth]{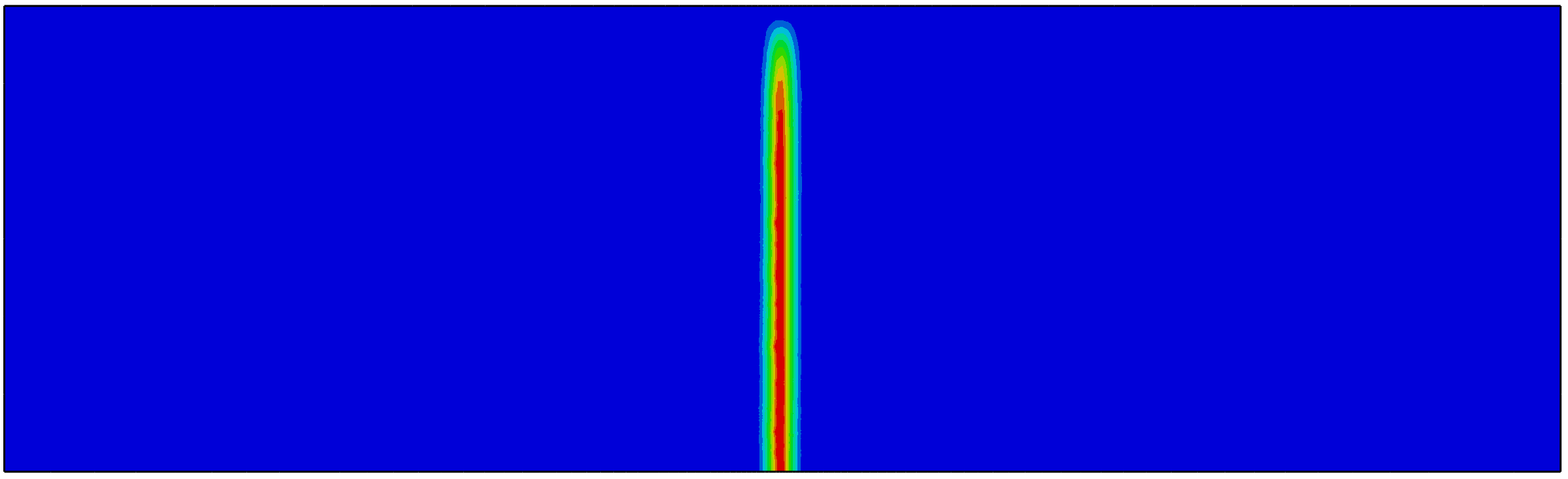}
        \caption{}
        \label{Fig:3point-phi}
    \end{subfigure}\vspace{0.02\textwidth}
    \begin{subfigure}[H]{0.3\textwidth}
        \includegraphics[width=\textwidth]{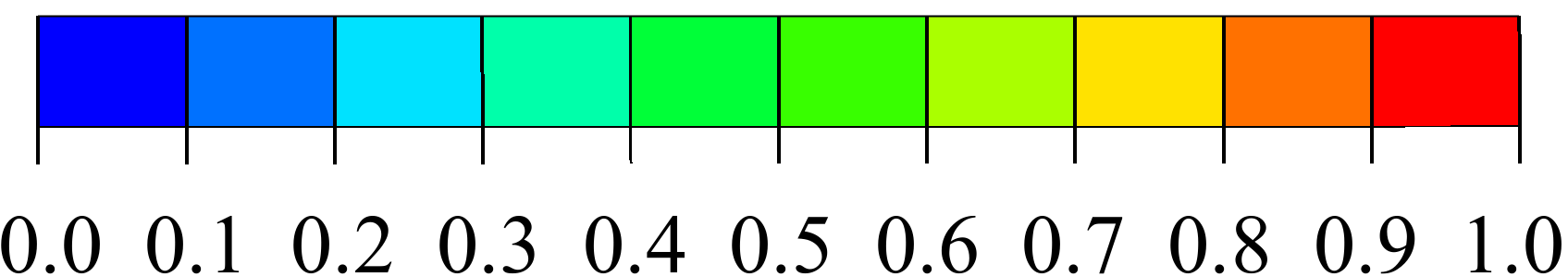}
    \end{subfigure}
    \begin{tikzpicture}[node distance=1cm]
    \node (cap) {{\Large $\phi$}};
    \end{tikzpicture}
    \caption{Three-point bending test: (a) geometry, dimensions and boundary conditions, (b) finite element mesh, and (c) phase field contour at the end of the analysis.}
    \label{Fig:3point}
\end{figure}
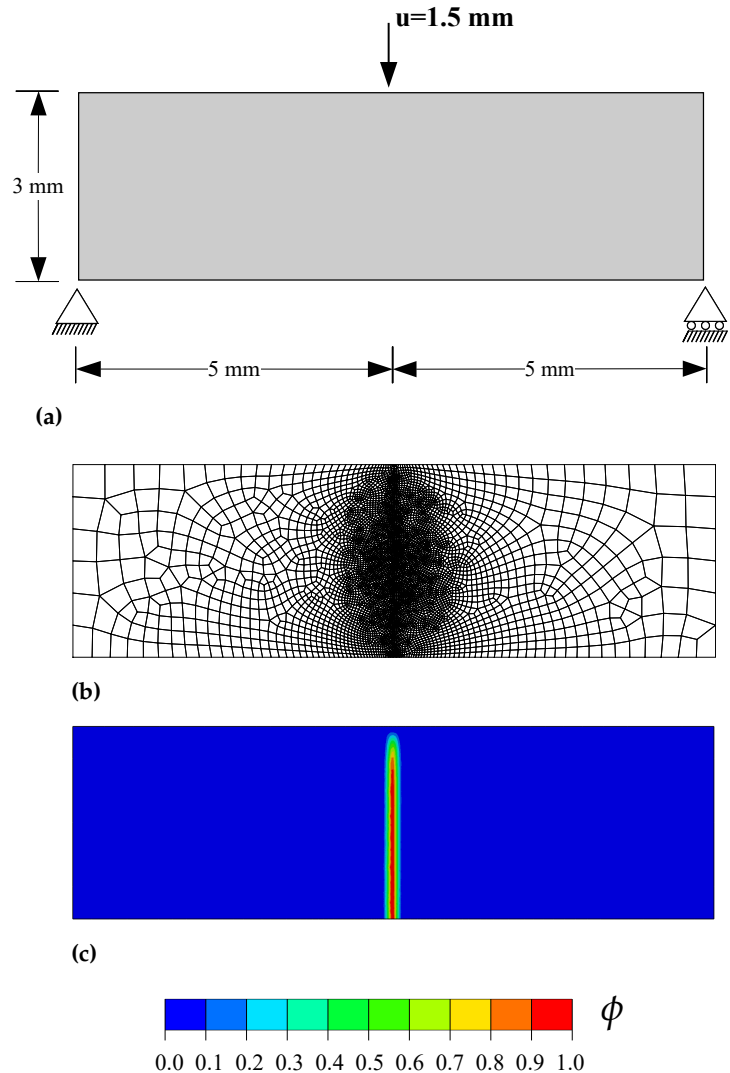

The geometry, dimensions and boundary conditions are shown in Fig. \ref{Fig:3point-geom}. A vertical displacement of 1.5 mm is applied at the top of the beam, at a horizontal distance of 5 mm to each of the supports. No initial crack is defined in the beam. Following Ref. \cite{Wells2001}, the mechanical behaviour of the beam is characterised by a Young's modulus of $E=100$ MPa and a Poisson's ratio of $\nu = 0$, while the fracture behaviour is characterised by a tensile strength of $f_t=1$ MPa and a toughness of $G_c=0.1$ N/mm. Recall that in the \texttt{PF-CZM} model the material strength is explicitly incorporated into the constitutive response and, as a consequence, results become largely insensitive to the choice of phase field length scale, which is here assumed to be $\ell = 0.1$ mm. The model is discretised using 4-node coupled temperature-displacement plane strain elements (CPE4T in Abaqus notation). As shown in Fig. \ref{Fig:3point-mesh}, the mesh is refined in the centre of the beam, where the crack is expected to nucleate and grow. The characteristic element is at least five times smaller than the phase field length scale and the total number of elements equals 5,820. Results are computed using the monolithic scheme. 

\begin{figure}[H]
    \centering
    \includegraphics[width=.6\textwidth]{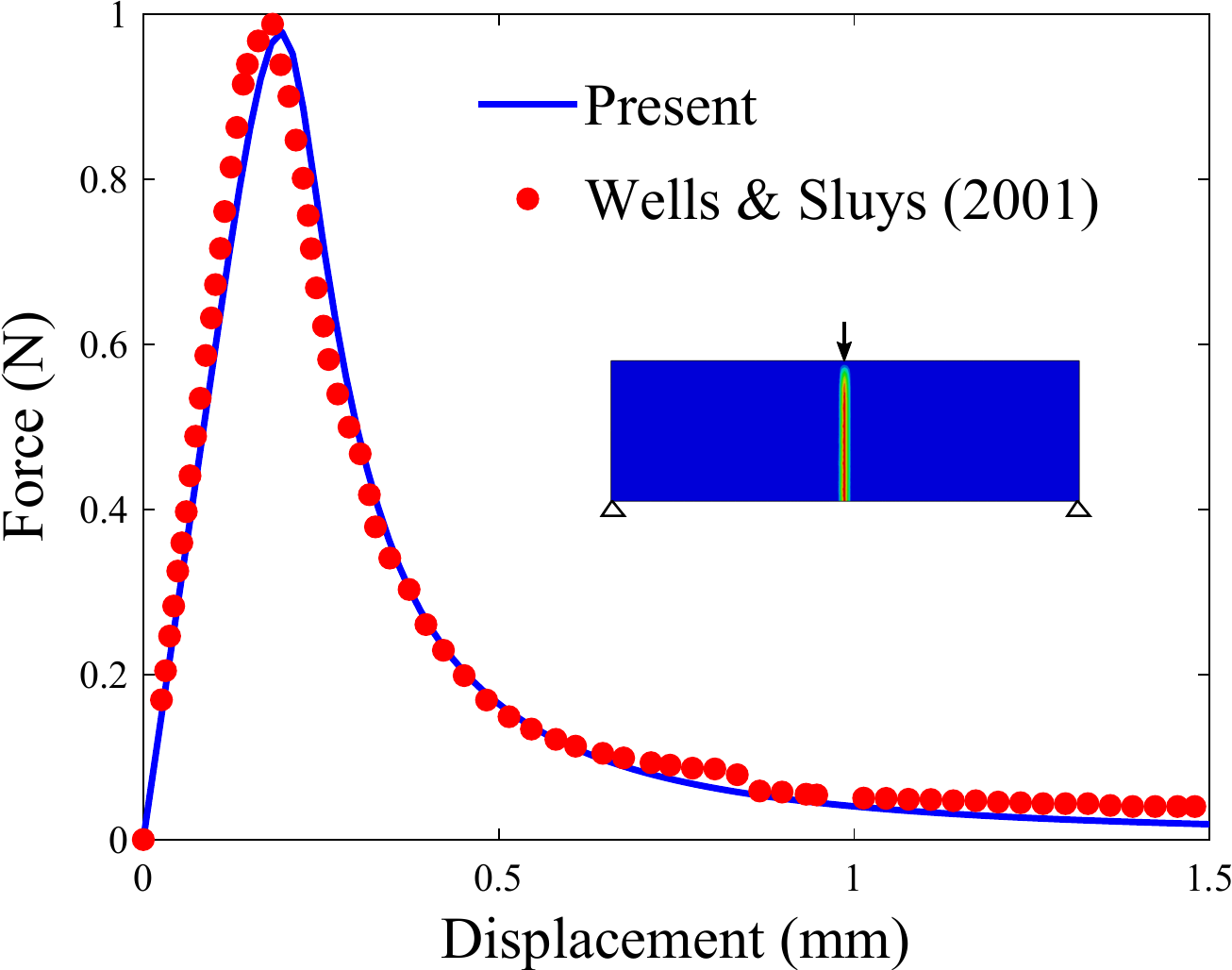}
    \caption{Three-point bending test: force versus displacement response. The results obtained in the present phase field are compared with the results computed by Wells and Sluys \cite{Wells2001} using an enriched cohesive zone model.}
    \label{Fig:3point-for-dis}
\end{figure}

In agreement with expectations and with the results by Wells and Sluys \cite{Wells2001}, a crack nucleates at the bottom of the beam, in the centre of the beam axis. The crack then propagates in a straight manner until reaching the top, as shown in Fig. \ref{Fig:3point-phi}. The resulting force versus displacement response reveals a quantitative agreement with the predictions by Wells and Sluys \cite{Wells2001} - see Fig. \ref{Fig:3point-for-dis}. 

\subsection{Mixed-mode fracture of a single-edge notched concrete beam}

\begin{figure}[H]
    \centering
    \includegraphics[width=.4\textwidth]{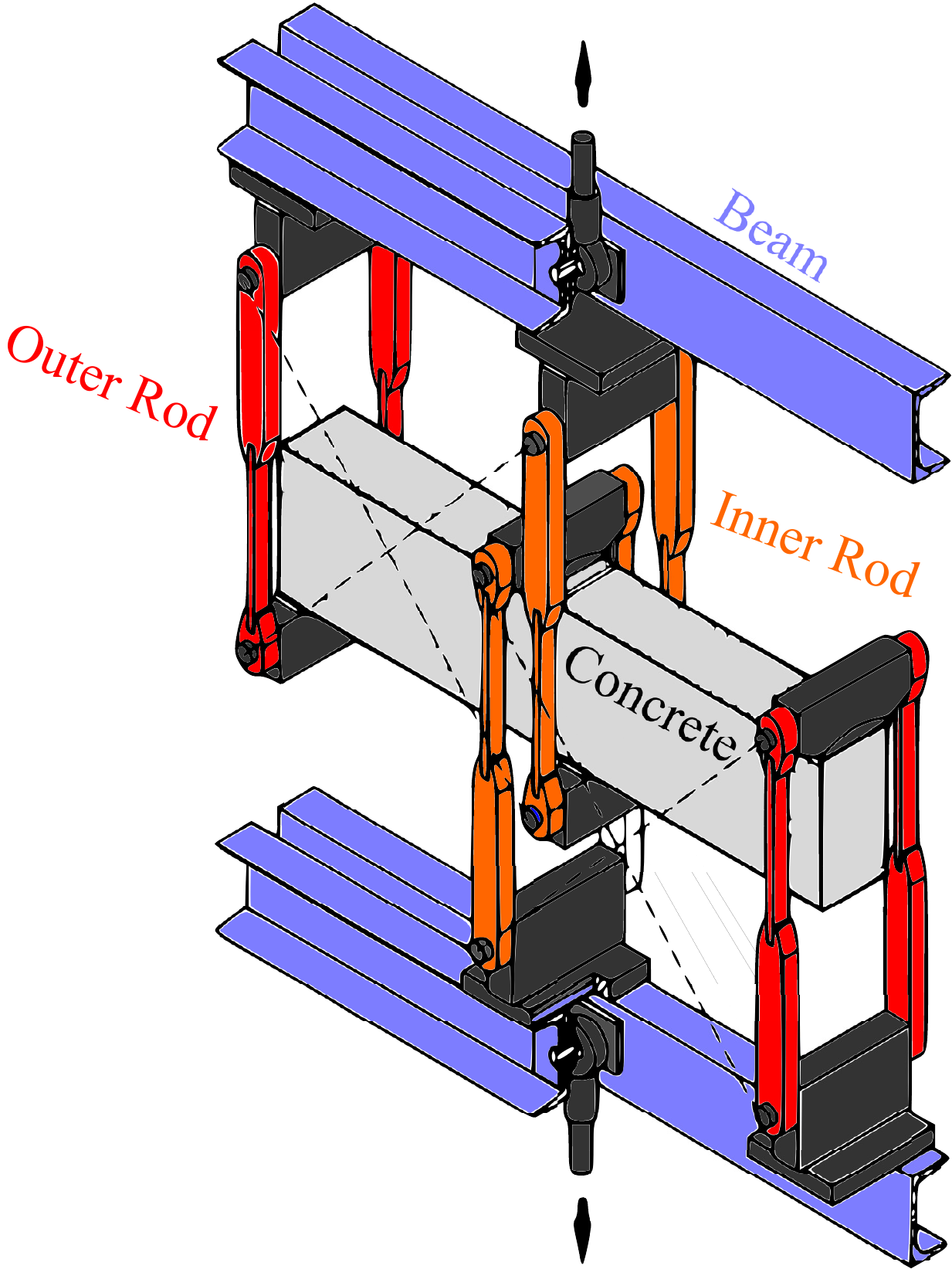}
    \caption{Mixed-mode fracture of a concrete beam: experimental testing configuration, following Ref. \cite{Schalangen1993}.}
    \label{Fig:SEN-setup}
\end{figure}

We proceed to model the failure of a concrete beam containing a notch. The aim is to compare the predictions obtained with the \texttt{AT2} model with the experimental observations by Schalangen \cite{Schalangen1993}. Schalangen subjected a concrete beam to the loading configuration shown in Fig. \ref{Fig:SEN-setup}. The beam is supported at four locations, and each support is connected to a girder beam through a rod. The cross-sections of the outer rods are smaller than those of the inner rods, to ensure an equal elongation. The load is applied to the centre of the girder beams and then transferred through the rods to the concrete beam. The resulting fracture is stable and mixed-mode. 

\begin{figure}[H]
    \centering
    \begin{subfigure}[H]{0.5\textwidth}
    \hspace{-.4 cm}
        \includegraphics[width=\textwidth]{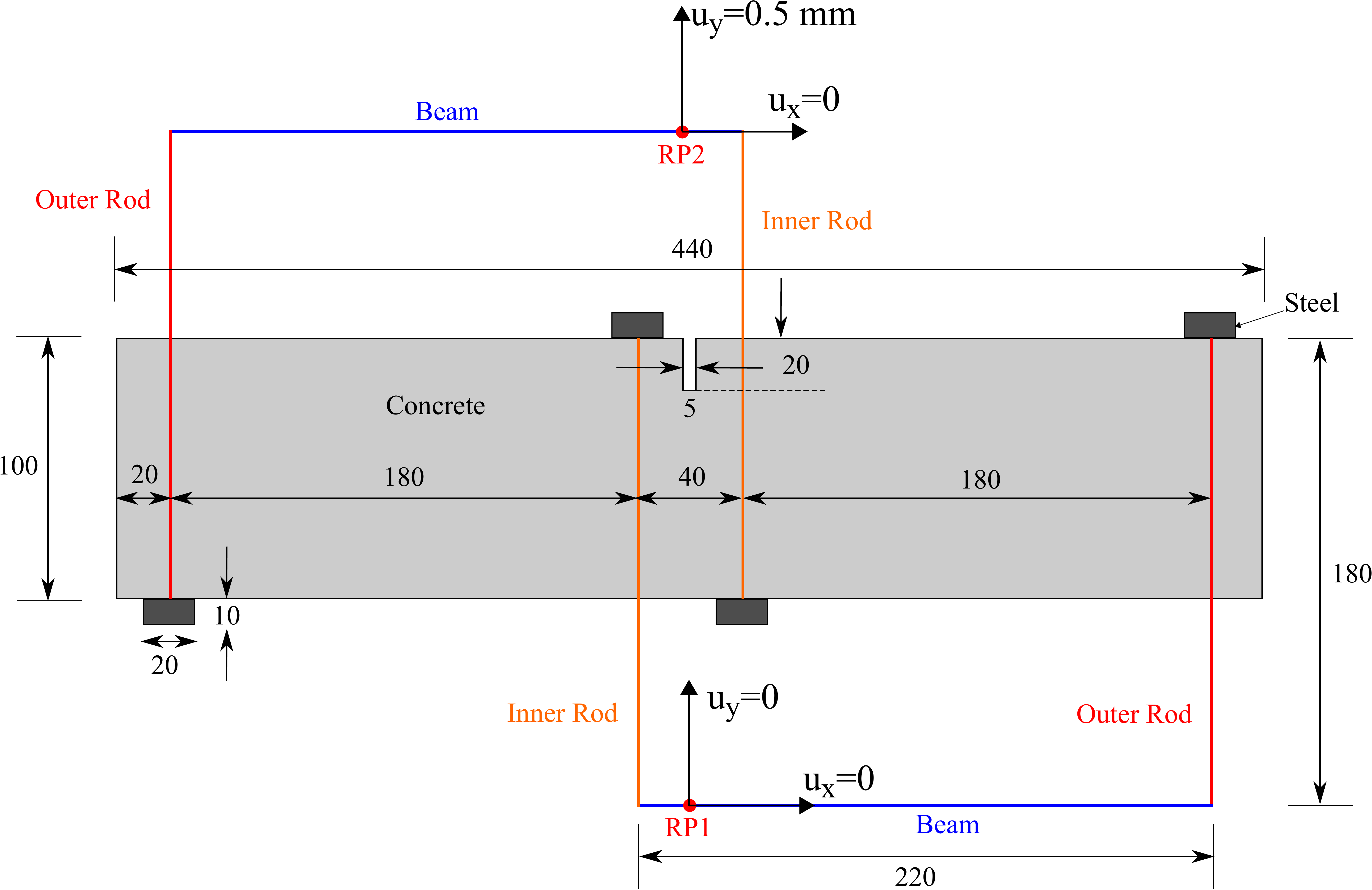}
        \caption{}
        \label{Fig:SEN-geom}
    \end{subfigure}\vspace{0.02\textwidth}
    \begin{subfigure}[H]{0.45\textwidth}
        \includegraphics[width=\textwidth]{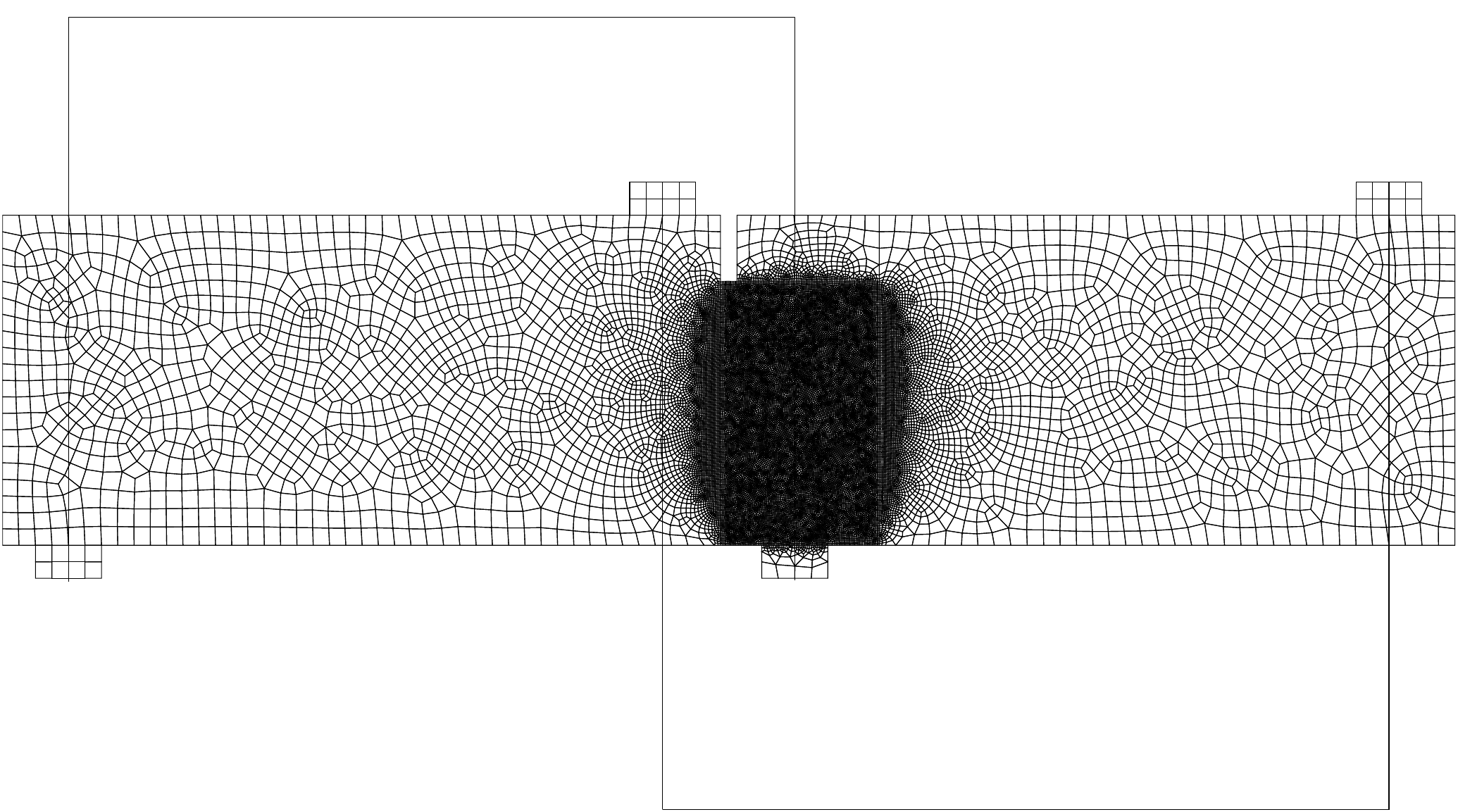}
        \caption{}
        \label{Fig:SEN-mesh}
    \end{subfigure}\vspace{0.02\textwidth}
    \caption{Mixed-mode fracture of a concrete beam: (a) geometry, dimensions (in mm) and boundary conditions, and (b) finite element mesh.}
    \label{Fig:SEN}
\end{figure}

The geometry and boundary conditions of our finite element model aim at mimicking the experimental configuration, see Fig. \ref{Fig:SEN-geom}. Two rigid beams are defined, tied to the reference points RP1 and RP2, where the boundary conditions are applied. Both girder beams can rotate around their reference points. The steel rods and supports are modelled and assigned a Young's modulus $E=210$ GPa and a Poisson's ratio equal to $\nu=0.3$. The cross-section of the inner rods equals 1,000 mm$^2$ while the cross-section of the outer rods is taken to be ten times smaller, in agreement with the experimental configuration. As shown in Fig. \ref{Fig:SEN-geom}, both horizontal and vertical displacements are constrained at the reference point RP1, while RP2 has its horizontal displacement constrained but is subjected to a vertical displacement of 0.5 mm. 

\begin{figure}[H]
    \centering
    \begin{subfigure}[H]{0.2\textwidth}
        \includegraphics[width=\textwidth]{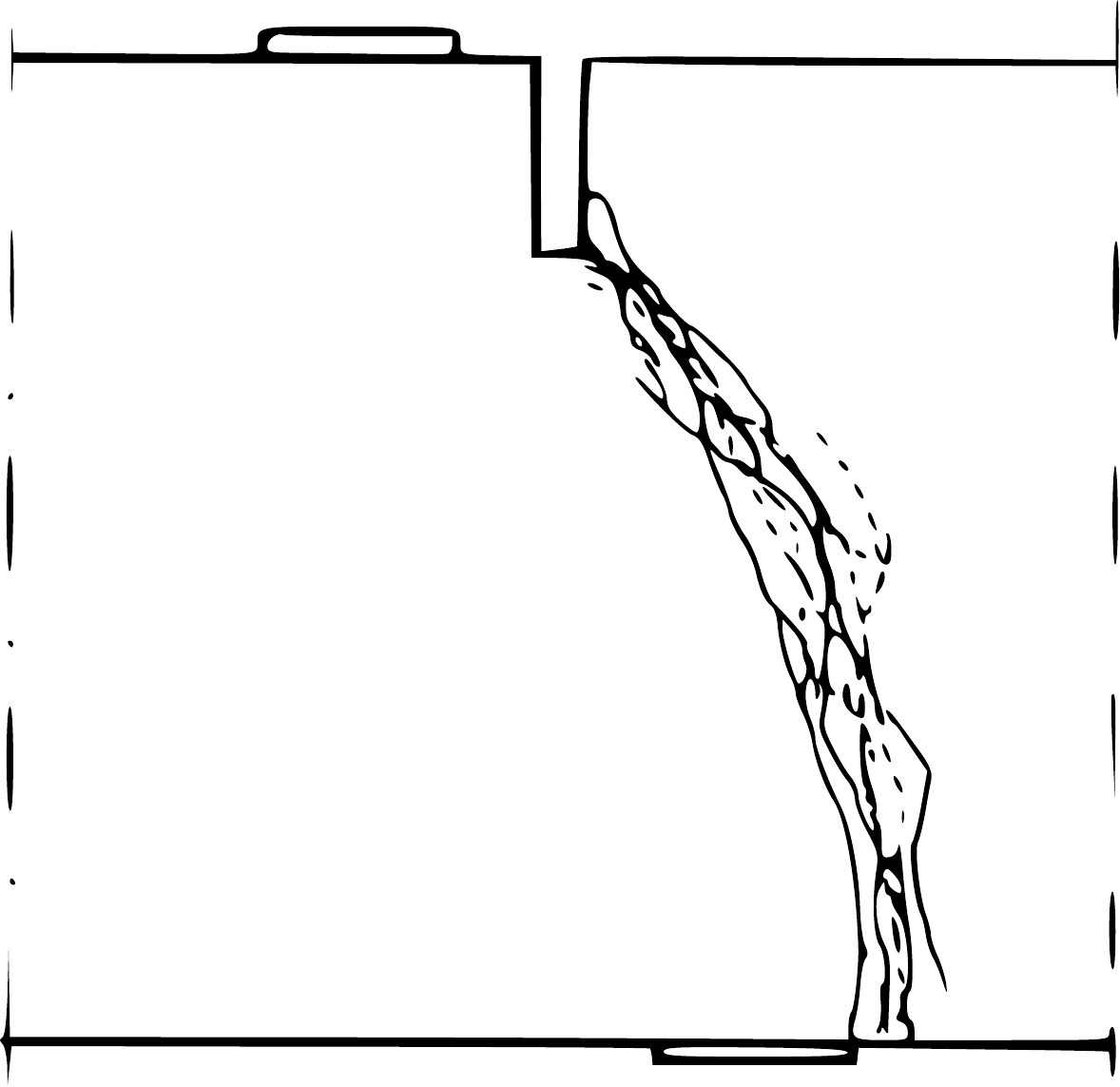}
        \caption{}
        \label{Fig:SEN-exper}
    \end{subfigure}\hspace{0.05\textwidth}
    \begin{subfigure}[H]{0.45\textwidth}
        \includegraphics[width=\textwidth]{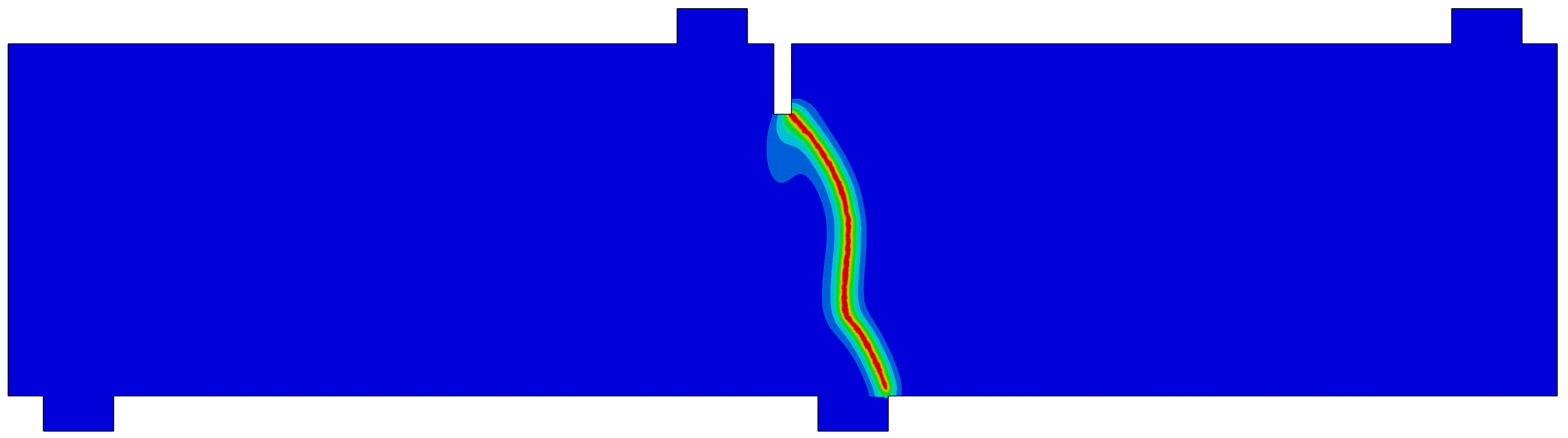}
        \caption{}
        \label{Fig:SEN-phi}
    \end{subfigure}
    \begin{subfigure}[H]{0.3\textwidth}
        \includegraphics[width=\textwidth]{legend-h-eps-converted-to.pdf}
    \end{subfigure}
    \begin{tikzpicture}[node distance=1cm]
    \node (cap) {{\Large $\phi$}};
    \end{tikzpicture}
    \caption{Mixed-mode fracture of a concrete beam: (a) Experimental crack patterns \cite{Schalangen1993}, and (b) predicted crack trajectory, as given by the phase field contour.}
    \label{Fig:SEN2}
\end{figure}

Fracture is simulated using the \texttt{AT2} model. To prevent failure of elements under compression, the strain energy density is divided into tensile and compressive parts employing the strain spectral decomposition proposed by Miehe \textit{et al.} \cite{Miehe2010a}, using the anisotropic formulation (\ref{eq:Anisotropic}). The material properties of the concrete beam are taken to be: Young's modulus $E=35$ GPa, Poisson's ratio $\nu=0.2$, and toughness $G_c=0.1$ N/mm. The phase field length scale is assumed to be equal to $\ell=2$ mm and, consequently, the characteristic size of the elements along the potential crack propagation region equals 0.5 mm (see Fig. \ref{Fig:SEN-mesh}). The rods are modelled using truss elements, while the concrete beam is discretised with a total of 28,265 linear quadrilateral coupled temperature-displacement plane strain elements. The results obtained are presented in Fig. \ref{Fig:SEN2}. Both experimental (Fig. \ref{Fig:SEN-exper}) and numerical (Fig. \ref{Fig:SEN-phi}) results are shown. A very good agreement can be observed, with the crack initiating in both cases at the right corner of the notch and deflecting, following a very similar trajectory, towards the right side of the bottom support. 

\subsection{Notched plate with an eccentric hole}

\begin{figure}[H]
    \centering
    \begin{subfigure}[H]{0.3\textwidth}
    \hspace{-.9 cm}
        \includegraphics[width=\textwidth]{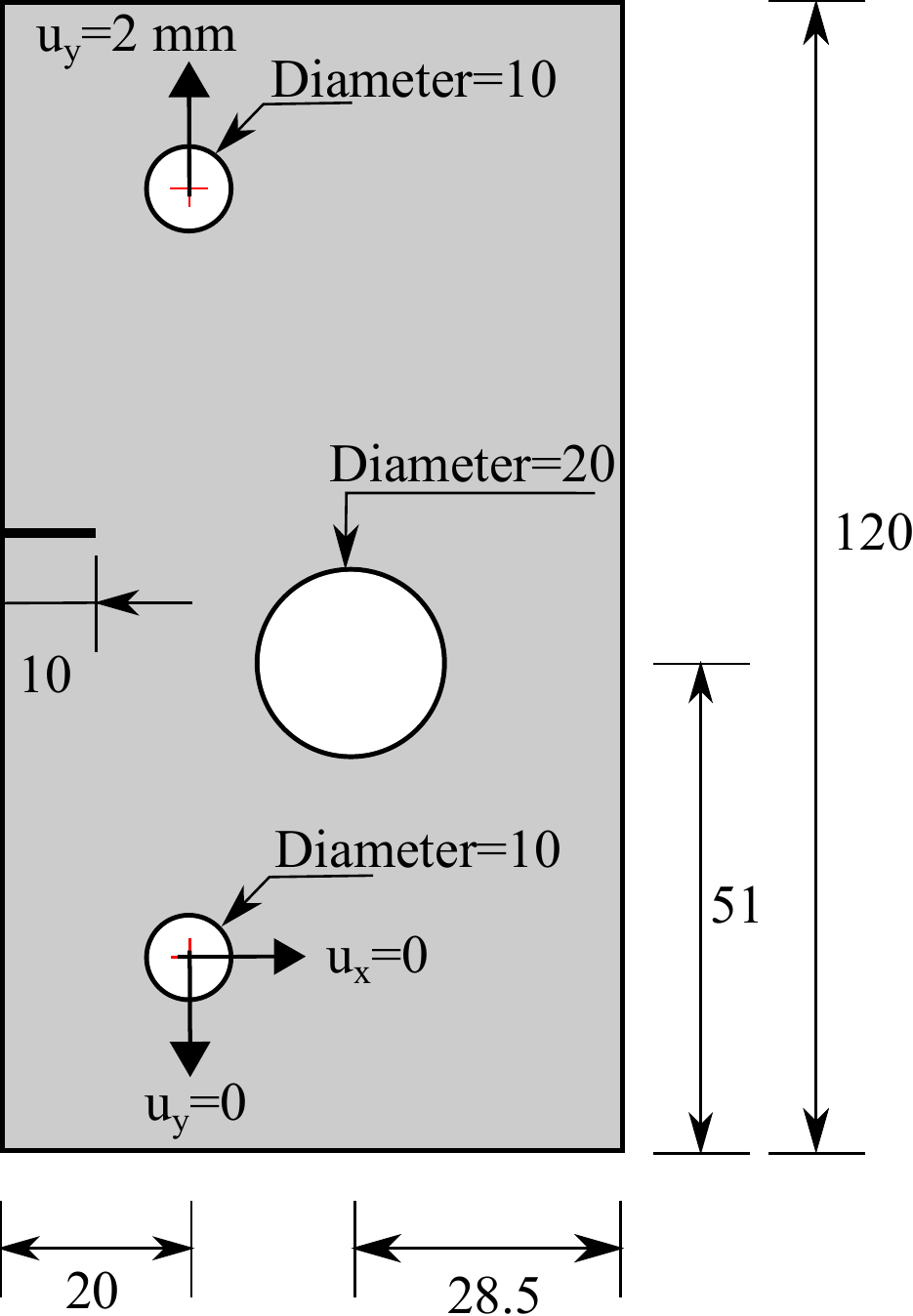}
        \caption{}
        \label{Fig:Notched plate-geo}
    \end{subfigure}\vspace{0.02\textwidth}
    \begin{subfigure}[H]{0.22\textwidth}
        \includegraphics[width=\textwidth]{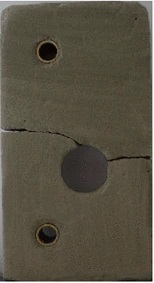}
        \caption{}
        \label{Fig:Notched plate-exper}
    \end{subfigure}\vspace{0.02\textwidth}
    \begin{subfigure}[H]{0.23\textwidth}
        \includegraphics[width=\textwidth]{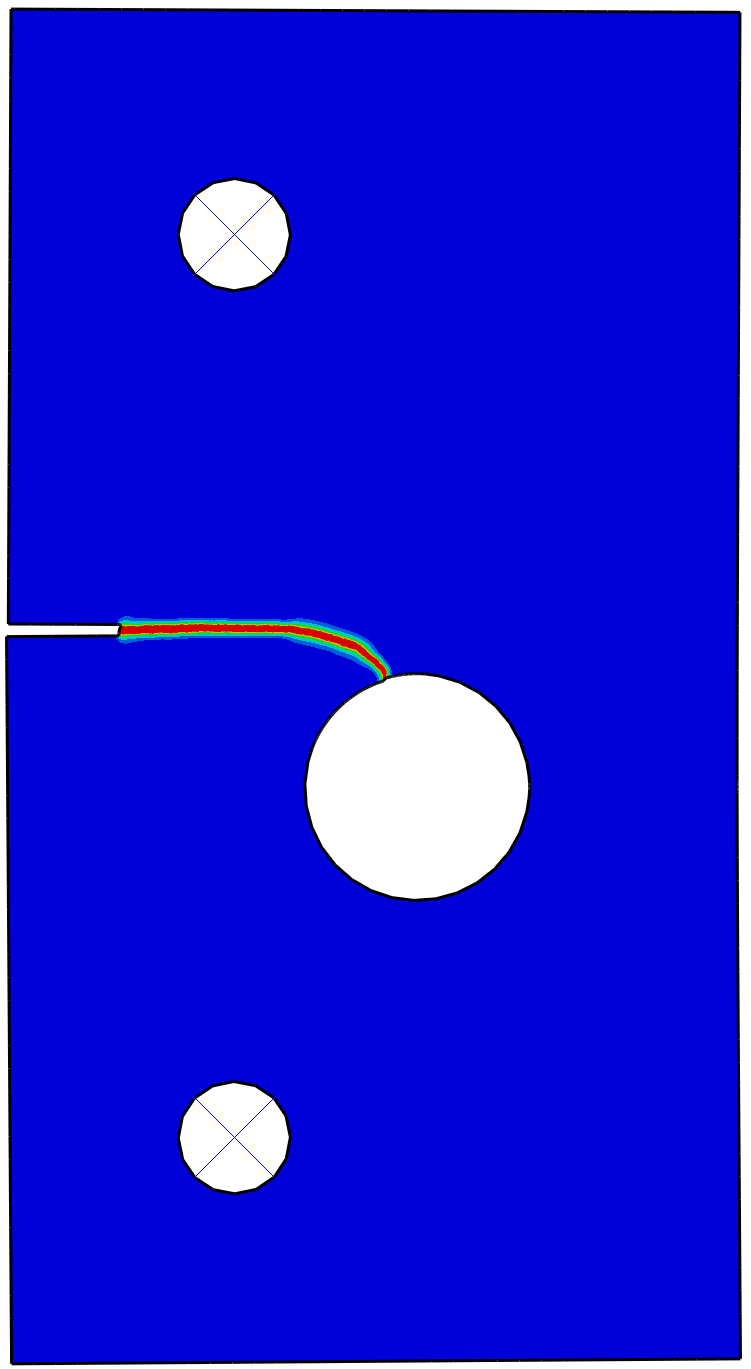}
        \caption{}
        \label{Fig:Notched plate-phi-40}
    \end{subfigure}\hspace{0.12\textwidth}
        \begin{subfigure}[H]{0.23\textwidth}
        \includegraphics[width=\textwidth]{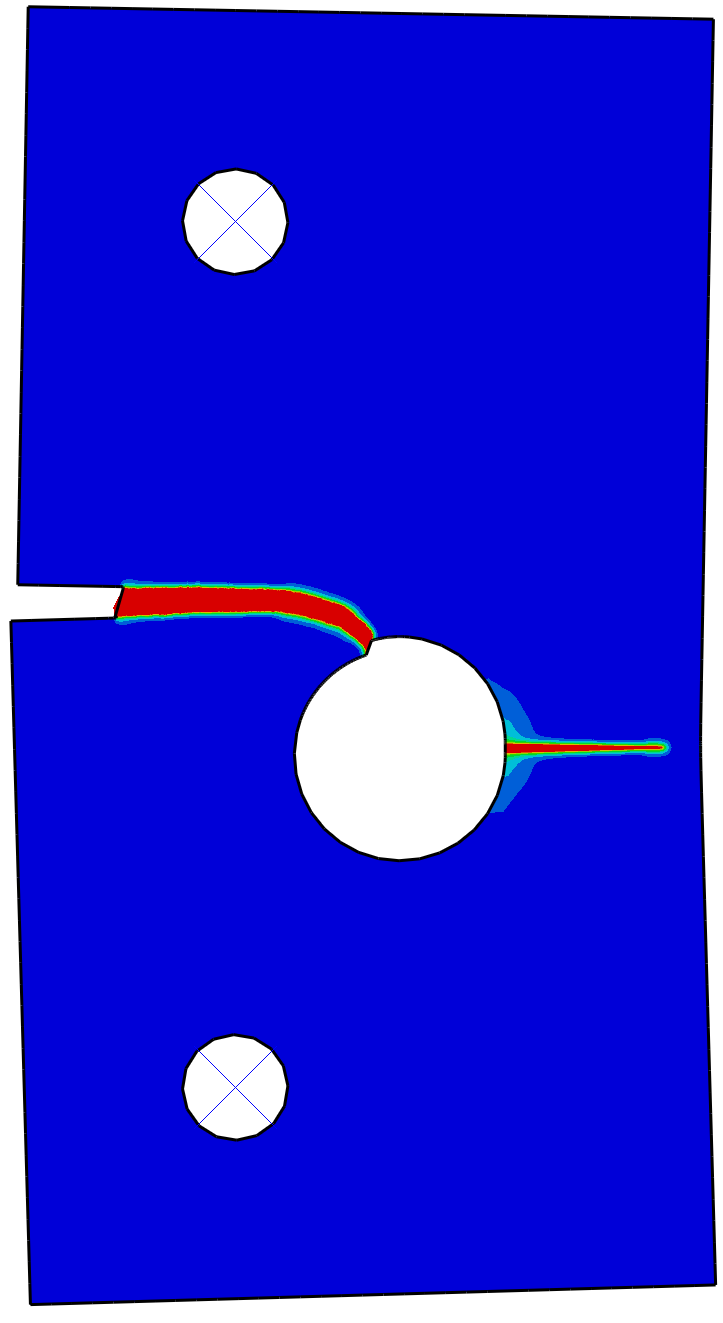}
        \caption{}
        \label{Fig:Notched plate-phi-200}
    \end{subfigure}\vspace{0.02\textwidth}
    \begin{subfigure}[H]{0.25\textwidth}
        \includegraphics[width=\textwidth]{legend-h-eps-converted-to.pdf}
    \end{subfigure}
    \begin{tikzpicture}[node distance=1cm]
    \node (cap) {{\Large $\phi$}};
    \end{tikzpicture}
    \caption{Notched plate with an eccentric hole: (a) geometry, dimensions (in mm) and boundary conditions, (b) experimental observation \cite{Ambati2015}, and predicted phase field $\phi$ contours at (c) $u=0.4$ mm and (d) $u=2$ mm.}
    \label{Fig:Notched plate}
\end{figure}

In this case study, we demonstrate the capabilities of the framework in capturing the interaction of cracks with other defects, and in predicting crack nucleation from arbitrary sites. This is achieved by using the monolithic scheme and without observing convergence issues. Specifically, we chose to model the failure of a mortar plate, which has been experimentally and numerically investigated by Ambati \textit{et al.} \cite{Ambati2015}. As shown in Fig. \ref{Fig:Notched plate-geo}, the plate contains a 10 mm notch and an eccentric hole of 10 mm radius. Mimicking the experimental setup, the plate contains two loading pin holes; the bottom one is fixed in both vertical and horizontal directions, while a vertical displacement of 2 mm is applied to the top one. The material properties are $E=5,982$ MPa, $\nu=0.22$, $\ell=0.25$ mm and $G_c=2.28$ N/mm. The \texttt{AT2} phase field model is considered, with no split applied to the strain energy density. We discretise the plate with 56,252 linear plane stress coupled displacement-thermal elements (CPS4T, in Abaqus notation). The characteristic element length in the regions surrounding the notch and the hole is five times smaller than the phase field length scale.

\begin{figure}[H]
    \centering
    \includegraphics[width=.7\textwidth]{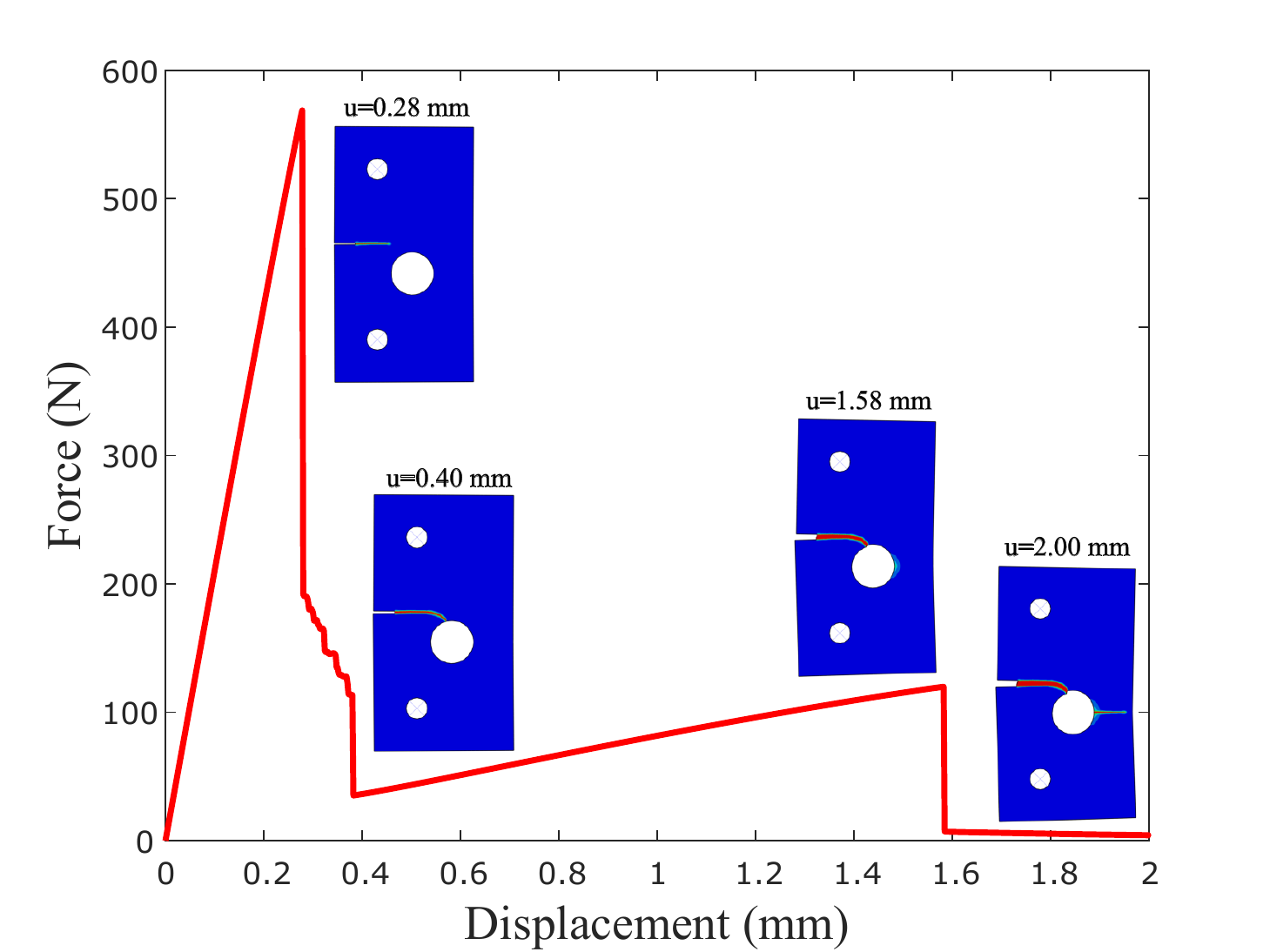}
    \caption{Notched plate with an eccentric hole: force versus displacement curve, with several snapshots of several cracking events superimposed.}
    \label{Fig:holed-for-dis}
\end{figure}

The results obtained, in terms of the crack trajectory, are shown in Fig. \ref{Fig:Notched plate}. A very good agreement with the experimental observations is attained (Fig. \ref{Fig:Notched plate-exper}). As shown in Fig. \ref{Fig:Notched plate-phi-40}, the crack starts from the notch tip and deflects towards the hole. The location of the point of interaction between the hole and the crack originating from the notch appears to be the same for experiments and simulations. Upon increasing the applied load, a new crack eventually nucleates from the right side of the hole, and propagates until reaching the end of the plate. The resulting force versus displacement response is shown in Fig. \ref{Fig:holed-for-dis}, where various images of the crack path have been superimposed to facilitate interpretation. The curve exhibits a linear behaviour until crack nucleation occurs ($u \approx 0.28$ mm), when a sudden drop in the load carrying capacity is observed. The interaction between the crack and the hole induces mixed-mode conditions and crack deflection, which is reflected in the force versus displacement curve. Once the crack has reached the hole, the applied displacement can be further increased without a drop in the load. This is observed until the nucleation of the second crack, which leads to the complete failure of the plate. 

\subsection{3D analysis of cracking due to the contact interaction between two gears}

\begin{figure}[H]
    \centering
    \includegraphics[width=.45\textwidth]{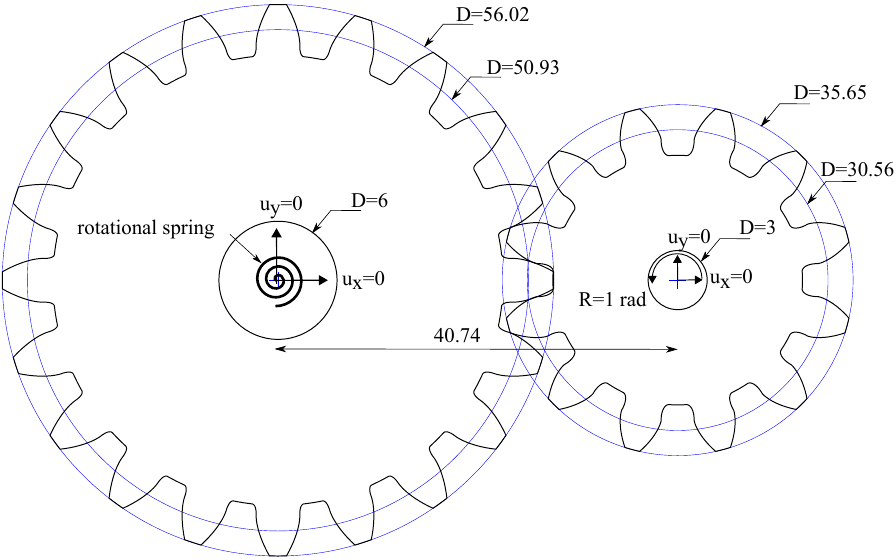}
    \caption{Cracking in interacting gears: geometry, dimensions (in mm) and boundary conditions}
    \label{Fig:Gear-geom}
\end{figure}

Finally, we proceed to showcase the abilities of the model in simulating complex 3D boundary value problems, involving advanced features such as contact. It should be emphasised that the same subroutine is used for both 2D and 3D analyses as the implementation is conducted at the integration point level. We chose to simulate the nucleation and growth of cracks in the teeth of two interacting gears, a problem of important technological relevance. The geometries of the two gears are shown in Fig. \ref{Fig:Gear-geom}, with dimensions given in mm. The circular pitch equals 8 mm, the pressure angle is 20$^\circ$ and both the clearance and the backlash equal 0.05 mm. Both gears have a thickness of 3 mm. The boundary conditions are also depicted in Fig. \ref{Fig:Gear-geom}. The inner hole of each gear is tied to the gear centre point. The centre of the small, right gear is subjected to a rotation of 1 radian, while a linear rotational spring is considered at the centre of the large, left gear. The stiffness of the rotational spring is $7 \times 10^6$ N$\cdot$mm/rad. 

\begin{figure}[H]
    \centering
    \begin{subfigure}[H]{0.45\textwidth}
        \includegraphics[width=\textwidth]{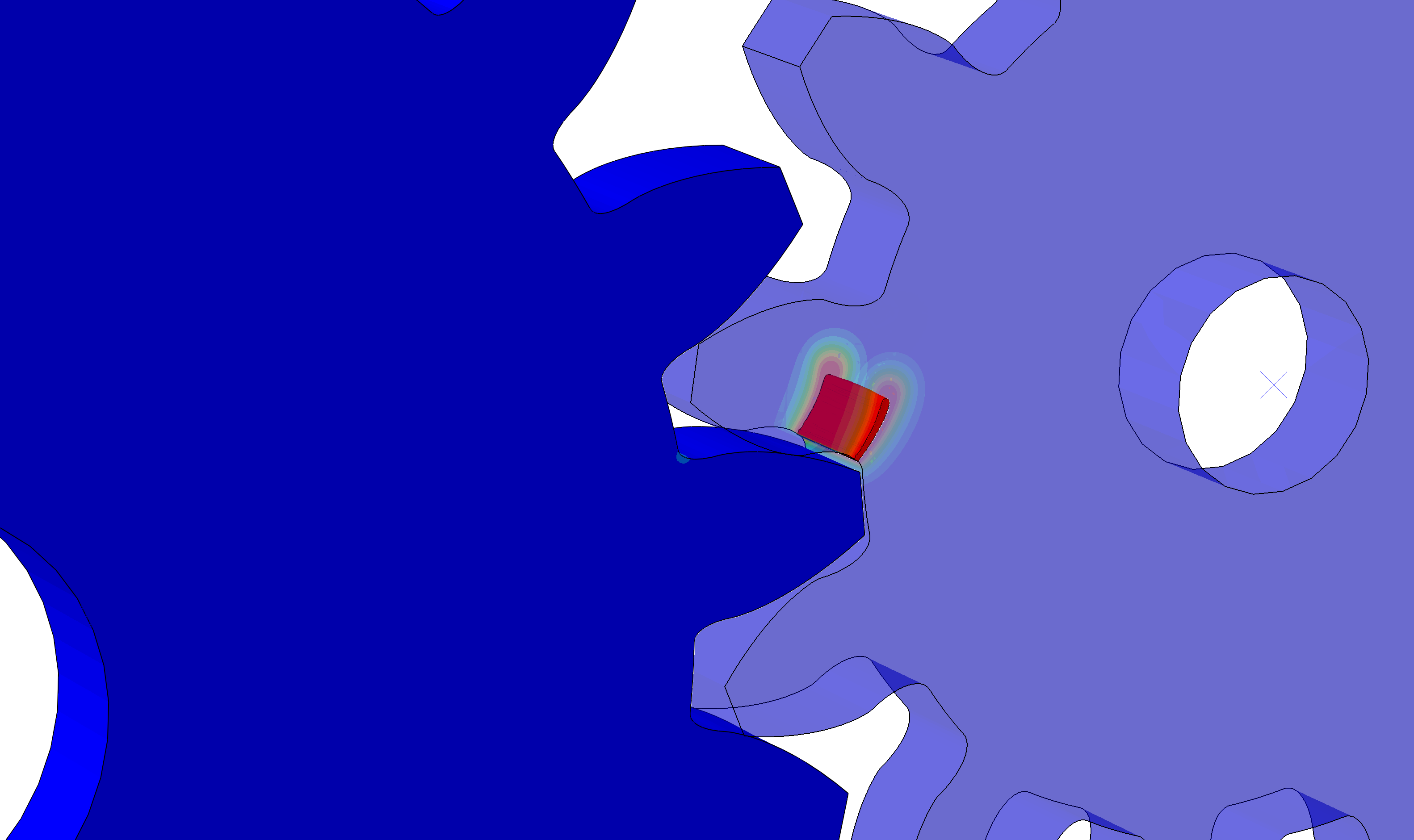}
        \caption{}
        \label{Fig:GearA}
    \end{subfigure}\vspace{.5 cm}
    \begin{subfigure}[H]{\textwidth}
        \includegraphics[width=.7\textwidth]{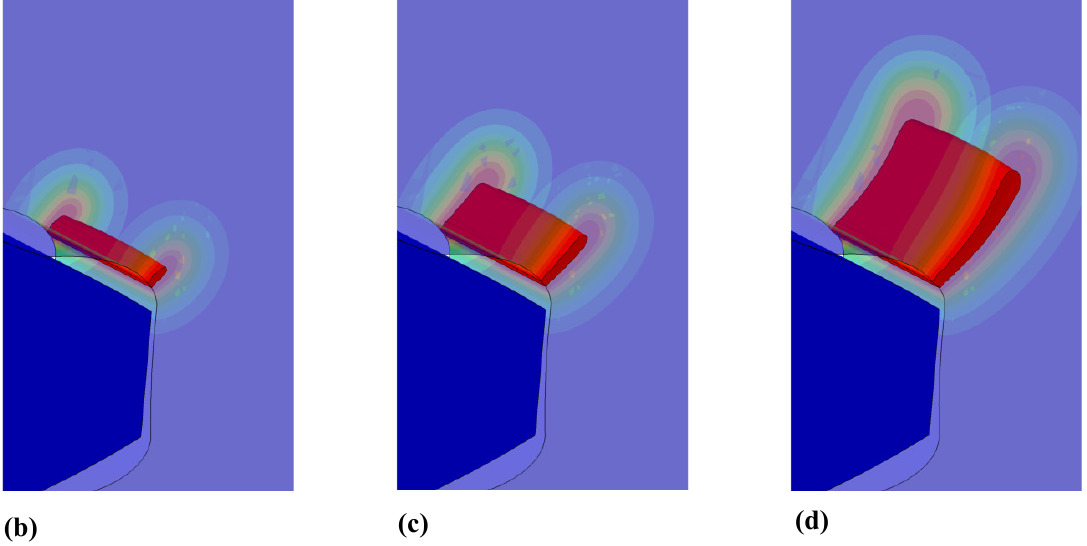}
        \label{Fig:GearA}
    \end{subfigure}
    \begin{subfigure}[H]{0.3\textwidth}
        \includegraphics[width=\textwidth]{legend-h-eps-converted-to.pdf}
    \end{subfigure}
    \begin{tikzpicture}[node distance=1cm]
    \node (cap) {{\Large $\phi$}};
    \end{tikzpicture}
    \caption{Cracking in interacting gears: phase field contours, (a) overall view at an advanced stage of cracking, and detail at (b) $0.028 +2 \times 10^{-7}$ rad, (c) $0.028 +5 \times 10^{-7}$ rad and (d) $0.028 +9 \times 10^{-7}$ rad.}
    \label{Fig:Gear}
\end{figure}

The modelling requires a non-linear geometrical analysis and the use of a contact algorithm to simulate the interaction between the gear teeth. Frictionless contact is assumed for the tangential contact behavior, which is enforced by making the Lagrangian multiplier equal to zero. The normal contact behavior is considered as a hard contact with a surface-to-surface interaction. The penetration of the slave surface into the master surface is minimised under hard contact conditions. The normal contact constraint is enforced through a Lagrangian multiplier. The material properties read $E=210$ GPa, $\nu = 0.3$, $\ell = 0.25$ mm, and $G_c= 2.7$ N/mm. Fracture is predicted using the \texttt{AT1} model and no split is used for the strain energy density. The model is discretised with more than 120,000 three-dimensional coupled temperature-displacement brick elements. The results obtained are shown in Fig. \ref{Fig:Gear}, in terms of phase field $\phi$ contours. Cracking initiates from the root of one of the teeth from the smaller gear and propagates towards the opposite root until the rupture of the gear teeth.

\section{Conclusions}
\label{Sec:Conclusions}

We have presented a unified Abaqus implementation of the phase field fracture method. Unlike previous works, our implementation requires only one user subroutine, of the user material type (UMAT). This enables avoiding the use of user elements, with the associated complications in pre and post-processing, as well as exploiting the majority of Abaqus' in-built features. The implementation is compact, requiring only 33 lines of code in its simpler form, and can be used indistinctly for 2D and 3D problems. It is also robust, as both staggered and monolithic solution schemes have been incorporated. Moreover, the implementation can accommodate any constitutive choice of phase field model. We present a unified theoretical framework that resembles the code, and particularise it to three of the most widely used phase field models: \texttt{AT1}, \texttt{AT2} and \texttt{PF-CZM}. In addition, several strain energy splits are considered, in the framework of both hybrid and anisotropic formulations.\\

We have demonstrated the robustness and capabilities of the framework presented by addressing several boundary value problems of particular interest. First, we showed that the \texttt{PF-CZM} version leads to an excellent agreement with the enriched cohesive zone model analysis by Wells and Sluys \cite{Wells2001} of crack nucleation and growth in a beam subjected to three-point bending. Secondly, we validated the crack trajectories predicted by the \texttt{AT2} model with the experimental observations by Schalangen \cite{Schalangen1993} on a concrete beam exhibiting mixed-mode fracture. Thirdly, we simulated the failure of a mortar plate with an eccentric hole to showcase the capabilities of the framework in capturing the interaction between cracks and other defects, as well as the nucleation of secondary cracks. The simulations agree qualitatively and quantitatively with the results obtained by Ambati \textit{et al.} \cite{Ambati2015}. Finally, we used the \texttt{AT1} version to model cracking due to the interaction between gears to showcase the capabilities of the model in dealing with 3D problems incorporating complex computational features, such as contact and geometric non-linearity. The codes developed have been made freely available, with examples and documentation, at \url{www.empaneda.com/codes}.

\vspace{6pt} 



\authorcontributions{Conceptualization, E. Mart\'{\i}nez-Pa\~neda, Y. Navidtehrani, C. Beteg\'{o}n; methodology, Y. Navidtehrani; software, E. Mart\'{\i}nez-Pa\~neda, Y. Navidtehrani; validation, E. Mart\'{\i}nez-Pa\~neda, Y. Navidtehrani; data curation, Y. Navidtehrani; writing---original draft preparation, E. Mart\'{\i}nez-Pa\~neda; writing---review and editing, E. Mart\'{\i}nez-Pa\~neda, Y. Navidtehrani, C. Beteg\'{o}n; project administration, E. Mart\'{\i}nez-Pa\~neda, C. Beteg\'{o}n; funding acquisition, E. Mart\'{\i}nez-Pa\~neda, C. Beteg\'{o}n. All authors have read and agreed to the published version of the manuscript.}

\funding{The authors would like to acknowledge financial support from the Ministry of Science, Innovation and Universities of Spain through grant PGC2018-099695-B-I00. E. Mart\'{\i}nez-Pa\~neda additionally acknowledges financial support from the Royal Commission for the 1851 Exhibition (RF496/2018).}



\conflictsofinterest{The authors declare no conflict of interest.} 




\appendixtitles{yes} 
\appendix
\setcounter{section}{0}
\setcounter{equation}{0}
\section{Weak formulation and finite element implementation}
\label{Sec:AppendixFEM}

The heat transfer analogy enables implementing the phase field fracture method in Abaqus using only an integration point level user subroutine. Thus, the definition of the element stiffness matrix $\bm{K}^e$ and the element residual vector $\bm{R}^e$ are carried out by Abaqus internally. However, both are provided here for the sake of completeness. Consider the principle of virtual work presented in Section \ref{Sec:Theory}. Decoupling the deformation and fracture problems, the weak form reads,
\begin{equation}
  \int_\Omega \Big\{ \big[ {g(\phi)} + \kappa \big] \bm{\sigma}_0 : \delta \bm{\varepsilon} \Big\} \text{d} V = 0   \, .
\end{equation}
\begin{equation}
\int_{\Omega} \left\{ {g^{\prime}(\phi)\delta \phi} \, \mathcal{H} +
        \frac{1}{2c_w} G_c \left[ \frac{1}{2 \ell} {w^{\prime}(\phi)} \delta \phi - \ell \nabla \phi  \nabla \delta \phi \right] \right\}  \, \mathrm{d}V = 0  \, .
\end{equation}

Now let us proceed with the finite element discretisation. Adopting Voig notation, the nodal variables for the displacement field $\mathbf{\hat{u}}$, and the phase field $\hat{\phi}$ are interpolated as: 
\begin{equation}\label{eq:Ndiscret}
\mathbf{u} = \sum_{i=1}^m \bm{N}_i \hat{\mathbf{u}}_i, \hspace{1cm} \phi =  \sum_{i=1}^m N_i \hat{\phi}_i \, ,
\end{equation}
\noindent where $N_i$ is the shape function associated with node $i$ and $\bm{N}_i$ is the shape function matrix, a diagonal matrix with $N_i$ in the diagonal terms. Also, $m$ is the total number of nodes per element such that $\hat{\mathbf{u}}_i=\left\{ u_x, \, u_y, \, u_z \right\}^T$ and $\hat{\phi}_i$ respectively denote the displacement and phase field at node $i$. Consequently, the associated gradient quantities can be discretised using the corresponding \textbf{B}-matrices, containing the derivative of the shape functions, such that:
\begin{equation}\label{eq:Bdiscret}
\bm{\varepsilon} = \sum\limits_{i=1}^m \bm{B}^{\bm{u}}_i \hat{\mathbf{u}}_i, \hspace{0.8cm}  \nabla\phi =  \sum\limits_{i=1}^m \mathbf{B}_i \hat{\phi}_i \, .
\end{equation}

Considering the discretisation (\ref{eq:Ndiscret})-(\ref{eq:Bdiscret}), we derive the residuals for each primal kinematic variable as:
\begin{align}
    & \mathbf{R}_i^\mathbf{u} = \int_\Omega \left\{\left[{g(\phi)}+ \kappa \right]\left(\bm{B}^\mathbf{u}_i\right)^T \bm{\sigma}_0 \right\} \, \text{d}V \, , \\
    & R_i^\phi = \int_\Omega \left\{ {g^{\prime}(\phi)}N_i \mathcal{H} + \frac{G_c}{2c_w \ell}  \left[\frac{w^{\prime}(\phi)}{2 } N_i + \ell^2 \,  \left( \mathbf{B}_i \right)^T \nabla\phi\right]\right\}dV \,.
\end{align}

Finally, the consistent tangent stiffness matrices $\bm{K}$ are obtained by differentiating the residuals with respect to the incremental nodal variables as follows:
\begin{align}
    & \bm{K}_{ij}^{\mathbf{u}} = \frac{\partial \bm{R}_{i}^{\bm{u}} }{\partial \bm{u}_{j} } = 
        \int_{\Omega} \left\{ \left[ {g(\phi)}+ \kappa \right] {(\bm{B}_{i}^{\bm{u}})}^{T} \bm{C}_0 \, \bm{B}_{j}^{\bm{u}} \right\} \, \text{d}V \, , \\
    & \bm{K}_{ij}^{\phi} = \frac{\partial R_{i}^{\phi} }{ \partial \phi_{j} } =  \int_{\Omega} \left\{ \left( {g''(\phi)} \mathcal{H} + \frac{G_{c}}{4 c_w \ell}  {w''(\phi)} \right) N_{i} N_{j} + \frac{G_{c} \ell}{2 c_w}   \, \mathbf{B}_i^T\mathbf{B}_j \right\} \, \text{d}V \, ,
\end{align}


\end{paracol}
\reftitle{References}


\externalbibliography{yes}
\bibliography{library}


\end{document}